\documentclass[cernpreprint, atlasdraft=true, texlive=2016, USenglish, texmf]{atlasdoc}
 
\usepackage[subcaption,backend=biber]{atlaspackage}
\usepackage{atlasbiblatex}
 
\usepackage{atlasphysics}

\newcommand{\LogCalRatio}{\log_{10}\left( E_{\text{HAD}} / E_{\text{EM}} \right)}
 
\graphicspath{{logos/}{figures/}}

\usepackage{multirow}

\AtlasTitle{Search for events with a pair of displaced vertices from long-lived neutral particles decaying into hadronic jets in the ATLAS muon spectrometer in $pp$ collisions at $\sqrt{s}=13$ TeV}

\AtlasAbstract{
A search for events with two displaced vertices from long-lived particles (LLP) pairs using data collected by the ATLAS detector at the LHC is presented. This analysis uses 139~fb$^{-1}$ of proton--proton collision data at $\sqrt{s} = 13$ TeV recorded in \mbox{2015--2018}. The search employs techniques for reconstructing vertices of LLPs decaying to jets in the muon spectrometer displaced between 3 m and 14 m with respect to the primary interaction vertex. The observed numbers of events are consistent with the expected background and limits for several benchmark signals are determined. For the Higgs boson with a mass of 125 GeV, the paper reports the first exclusion limits for branching fractions into neutral long-lived particles below 0.1\%, while branching fractions above 10\% are excluded at 95\% confidence level for LLP proper lifetimes ranging from 4 cm to 72.4 m. In addition, the paper present the first results for the decay of LLPs into into $t\overline{t}$ in the ATLAS muon spectrometer.}

\AtlasRefCode{EXOT-2019-24}
 
\PreprintIdNumber{CERN-EP-2021-195}

\arXivId{2202.00587}
 
\HepDataRecord{127764}
 
\AtlasJournal{Phys.\ Rev.\ D.}
\AtlasJournalRef{\PRD 106 (2022) 032005}
\AtlasDOI{10.1103/PhysRevD.106.032005}

\hypersetup{pdftitle={ATLAS document},pdfauthor={The ATLAS Collaboration}}
 
\begin{document}
 
\maketitle
 
\tableofcontents

\section{Introduction}
\label{sec:intro}
 
The discovery of the Higgs boson at the LHC completed the Standard Model (SM) of elementary particles and focused attention on the many central features of our universe that the SM does not address: dark matter, neutrino mass, matter--antimatter asymmetry, and the hierarchy problem (naturalness). Many beyond-the-SM (BSM) theoretical constructs that address these phenomena predict the existence of long-lived particles (LLPs), with macroscopic decay lengths.

Examples of these constructs include supersymmetric (SUSY) models such as mini split SUSY \cite{Arvanitaki:2012ps, ArkaniHamed:2012gw}, gauge-mediated SUSY breaking \cite{Giudice:1998bp}, $R$-parity-violating SUSY \cite{Barbier:2004ez, Csaki:2013jza} and Stealth SUSY \cite{Josh1,Josh2}; models addressing the hierarchy problem such as Neutral Naturalness \cite{Chacko:2015fbc,Burdman:2006tz, Cai:2008au, Chacko:2005pe} and Hidden Valley~\cite{Strassler:2006im,Strassler:2006ri}; models addressing dark matter \cite{Baumgart:2009tn, Kaplan:2009ag, Chan:2011aa, Dienes:2011ja, Dienes:2012yz}, and the \mbox{matter--antimatter} asymmetry of the universe \cite{CuiSundrum:2012,Cui:2013,Cui:2014twa}; and models that generate neutrino masses \cite{Helo:2013esa,Batell:2016zod}.  Many of these theoretical models result in neutral LLPs, which may be produced in the proton--proton collisions at the LHC and decay back into SM particles far from the interaction point (IP). Such LLP decays can result in secondary decays significantly displaced from the IP that are usually referred to as displaced vertices (DVs).
 
Searches for LLPs decaying into final states containing jets have been carried out at LEP by DELPHI~\cite{LLPDelphi}, at the Tevatron by both the CDF~\cite{CDF} and D0~\cite{D0} collaborations, and at the LHC by the ATLAS, CMS and LHCb collaborations~\cite{HVAtlas,LHCb,HVAtlas8TeV,SUSY-2014-02,CMS,CalRatio, LHCb-PAPER-2016-065,LHCb-1612.00945,CMS-EXO-16-003,CMS-EXO-17-018, CMS-EXO-18-007, EXOT-2017-05, EXOT-2017-24, SUSY-2016-08, EXOT-2017-25,LHCb-2011.05263,LHCb-1612.00945,LHCb-1705.07332}. To date, no search has observed evidence of BSM, neutral LLPs.
 
This search focuses on DVs occurring in the ATLAS muon spectrometer (MS) and uses 139 fb$^{-1}$ of 13~\TeV\ $pp$ collision data. The event selection criteria and vertex reconstruction algorithms choose candidate events with two MS DVs. The results are interpreted in terms of a scalar portal model, but the analysis may be sensitive to additional models. These results supersede those of a previous search~\cite{EXOT-2017-05} for events with two MS DVs in a smaller 13~\TeV\ dataset. Furthermore, the current analysis uses an improved background estimation methodology, more advanced modeling of the trigger and vertex reconstruction in Monte Carlo (MC) simulation, as well as a refined signal efficiency extrapolation as a function of the LLP lifetime.

The paper describes the ATLAS detector in Section~\ref{sec:detector}, followed by an overview of the analysis strategy in Section~\ref{sec:strategy} and the theoretical models in Section~\ref{sec:models}. Details of the $pp$ collision data and MC simulation are provided in Section~\ref{sec:datamc}, while those of the specialized trigger and the purpose-built displaced vertex reconstruction algorithm follow in Section~\ref{sec:trigvtxreco}. The baseline event selection is described in Section~\ref{sec:baseselection}, while Section~\ref{sec:2MSVxSelection} describes the signal selection and Section~\ref{sec:Bkg2MSVx} describes the background estimation. The systematic uncertainties are discussed in Section~\ref{sec:syst}, and the results are presented in Section~\ref{sec:result}.

\section{ATLAS detector}
\label{sec:detector}
 
The ATLAS detector~\cite{PERF-2007-01}, which has nearly 4$\pi$ steradian coverage,\footnote{ATLAS uses a right-handed coordinate system with its origin at the nominal IP in the center of the detector and the $z$-axis along the beam pipe. The $x$-axis points from the IP to the center of the LHC ring, and the $y$-axis points upwards. Cylindrical coordinates $(r,\phi)$ are used in the transverse plane, where $\phi$ is the azimuth angle around the $z$-axis. The pseudorapidity is defined in terms of the polar angle $\theta$ as $\eta = - \ln \tan(\theta/2)$. Angular distance is measured in units of $\Delta R \equiv \sqrt{(\Delta\eta)^{2} + (\Delta\phi)^{2}}$.} is a multipurpose detector consisting of an inner tracking detector (ID) surrounded by a superconducting solenoid providing a \SI{2}{\tesla} axial magnetic field, electromagnetic and hadronic calorimeters, and a muon spectrometer (MS) based on three large air-core toroidal superconducting magnets, each with eight coils. The field integral of the toroids ranges between \num{2.0} and \SI{6.0}{\tesla\metre}
across most of the detector.

The ID covers the pseudorapidity range \(|\eta| < 2.5\). It consists of a silicon pixel detector, a silicon microstrip detector, and a straw-tube transition-radiation tracker.

The calorimeter system covers the pseudorapidity range $|\eta| < 4.9$. It consists of a high-granularity electromagnetic calorimeter (ECal) surrounded by a hadronic calorimeter (HCal). Within the region $|\eta|< 3.2$, the ECal comprises liquid-argon (LAr) barrel and endcap electromagnetic calorimeters with lead absorbers. An additional thin LAr presampler covering $|\eta| < 1.8$ is used to correct for energy loss in material upstream of the calorimeters. The ECal extends from 1.5~m to 2.0~m in $r$ in the barrel and from 3.6~m to 4.25~m in $|z|$ in the endcaps.
The HCal is a steel/scintillator-tile calorimeter that is segmented into three barrel structures within $|\eta| < 1.7$, and two copper/LAr hadronic calorimeters in the endcaps ($1.5 < |\eta| < 3.2$). The HCal covers the region from 2.25~m to 4.25~m in $r$ in the barrel (although the HCal active material extends only up to 3.9~m) and from 4.3~m to 6.05~m in $|z|$ in the endcaps. The solid angle coverage is completed with forward copper/LAr and tungsten/LAr calorimeter modules optimized for electromagnetic and hadronic measurements, respectively. Together the ECal and HCal have a thickness of 9.7~interaction lengths at $\eta=0$.
 
The MS comprises three stations of separate trigger and tracking chambers that measure the deflection of muons in a magnetic field generated by the air-core toroid magnets. The barrel and endcap chamber systems are subdivided into 16 sectors: 8 large sectors and 8 small sectors. For the barrel, the small sectors are inside each of the 8 magnet coils and the large sectors are between the coils. The endcap TGC chambers are arranged into 24 or 48 sectors depending on their $\eta$ and radial position.
Three stations of resistive-plate chambers (RPC) and thin-gap chambers (TGC) are used for triggering in the MS barrel and endcaps, respectively. The first two RPC stations, which are radially separated by 0.5~m, begin at a radius of either 7~m (large sectors) or 8~m (small sectors). The third station is located at a radius of either 9~m (large sectors) or 10~m (small sectors). In the endcaps, the first TGC station is located at $|z|=13$~m. The other two stations start at $|z|=14$~m and $|z|=14.5$~m, respectively. The muon trigger system covers the range $|\eta| < 2.4$.
The muon tracking chamber system covers the region $|\eta| < 2.7$ with three stations of monitored drift tubes (MDT), complemented by cathode-strip chambers (CSC) in the forward region. The MDT chambers consist of two multilayers separated by a distance ranging from 6.5~mm to 317~mm. Each multilayer consists of three or four layers of drift tubes. The individual drift tubes are 30~mm in diameter and have lengths of 2--5~m (barrel) and 2--6.5~m (endcaps) depending on the location of the chamber in the spectrometer.
In each multilayer, charged-particle track-segment reconstruction entails finding the line that is tangent to the drift circles. These single-multilayer segments are local measurements of the position and direction of the charged particle. Because of its design, the MDT measurement provides only a very coarse $\phi$ position of the track hit. In order to reconstruct the $\phi$ position and direction, the MDT measurements are combined with the $\phi$-coordinate measurements from the trigger chambers.
 
The ATLAS trigger and data acquisition system~\cite{TRIG-2016-01} consists of a hardware-based first-level (L1) trigger followed by a software-based high-level trigger (HLT) that reduces the rate of events selected for offline storage to 1~kHz.
 
The implementation of the L1 muon trigger logic is similar for the RPC and TGC systems. Each of the three stations of the RPC system and the two outermost stations of the TGC system consist of a doublet of independent detector layers.  The first TGC station contains a triplet of detector layers. The transverse momentum (\pT) of the muon candidate is measured by the L1 muon trigger, using different algorithms for low-\pT and high-\pT triggers. In the barrel, a low-\pT (${<}10$~\GeV) muon region-of-interest (RoI) is generated by requiring a coincidence of hits in at least three of the four layers of the two inner RPC stations. In the endcaps, the trigger requires hits in the two outer TGC stations. A high-$\pT$ muon RoI in the barrel requires additional hits in at least one of the two layers of the outer RPC station, while for the endcaps, additional hits in two of the three layers of the innermost TGC station are required.
The muon RoIs have an angular extent of $0.2\times0.2$ in $\Delta\eta \times \Delta\phi$ in the MS barrel and $0.1\times0.1$ in $\Delta\eta \times \Delta\phi$ in the MS endcaps.

An extensive software suite~\cite{ATL-SOFT-PUB-2021-001} is used in the reconstruction and analysis of real and simulated data, in detector operations, and in the trigger and data acquisition systems of the experiment.

\section{Analysis strategy}
\label{sec:strategy}
 
The analysis presented in this paper searches for events with two DVs in the MS. The benchmark models that motivate these strategies are discussed in detail in Section~\ref{sec:models}.
 
Candidate events are selected by the Muon RoI Cluster HLT trigger~\cite{MuonRoITrigger}, and a dedicated algorithm is used to reconstruct DVs in the MS~\cite{PERF-2013-01}.
Additional selection criteria, optimized by comparing expected signal events with background events, are used to maximize the analysis sensitivity.
 
The main background to LLPs decaying into hadronic jets in the MS originates from hadronic or electromagnetic showers not fully contained in the calorimeter volume (\emph{punch-through jets}) that result in tracks reconstructed in the MS.
Multijet events that result in vertices in the MS will often have ID tracks and calorimeter jets that point towards the MS DV. To reduce the acceptance of such fake vertices from multijet events, vertices are required to be isolated in $\eta$ and $\phi$ from ID tracks and calorimeter jets. Multijet events with mismeasured jets are another source of background as they will not be removed by the isolation criteria.

Additional background, referred to as \emph{noncollision background}, can be generated by electronic noise in the MDT and RPC/TGC chambers, by cosmic-ray muons, and by the LHC beam (beam-induced background~\cite{BIB}). This last contribution is composed of hadronic and electromagnetic showers caused by beam protons interacting with collimators or residual gas molecules inside the vacuum pipe.
 
To avoid unintended biasing of the results during the development of the analysis, the number of events observed in the signal region was not measured until the entire set of selection criteria were optimized and all the systematic uncertainties completely evaluated.

\section{Signal model}
\label{sec:models}
 
Although this analysis is sensitive to a large variety of models, this paper interprets the results in terms of a \emph{scalar portal} model~\cite{Strassler:2006ri},
where a Higgs boson with a mass of 125~\GeV\ or a lower/higher-mass scalar boson, both indicated by the symbol $\varPhi$, decays into two long-lived scalars.
 
A standard way of introducing long-lived, neutral particles to the SM is through a hidden sector that weakly couples to the SM. Scalar portals, where a 125~\GeV\ Higgs boson or a lower/higher-mass scalar boson mixes weakly with a hidden-sector scalar field, can result in pair production of long-lived hidden-sector scalars or pseudoscalars that carry no SM quantum numbers.
An indirect upper limit on the BSM Higgs boson decays of the order of 20\% can be derived from combined studies of the Higgs boson production and decay~\cite{HIGG-2018-57,CMS-HIG-17-023} under several assumptions, potentially allowing sizable branching fractions for decays into non-SM particles.
 
Long-lived scalars with a mass from 5 to 30~\GeV\ are also well motivated by naturalness models that are generic extensions of hidden-valley portal models~\cite{Uncolored}.
 
\begin{figure}[thbp]
\centering
\includegraphics[width=0.35\textwidth]{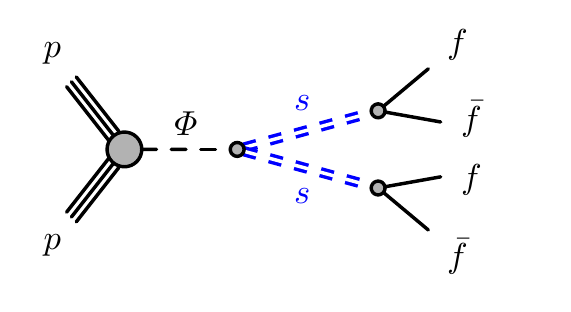}
\caption{
Diagram of the scalar portal model studied in this paper. The LLPs are represented by double lines and labeled $s$, and the final-state SM fermions are labeled as $f$~\cite{Strassler:2006ri}.}
\label{fig:HiggsFeyn}
\end{figure}

The mechanism for LLP production is shown in Figure~\ref{fig:HiggsFeyn}. Here, a scalar boson $\varPhi$ decays with some effective coupling into a pair of long-lived scalars, $s$. The scalars $s$ subsequently decay into SM particles.
In this model the couplings of the scalar $s$ to SM fermions are determined by the Higgs Yukawa coupling. Therefore, each long-lived scalar decays mainly into the heaviest fermion pair that is kinematically accessible. The branching fractions of $b\overline{b}$, $c\overline{c}$, and $\tau^{+}\tau^{-}$ decays depend on the mass of the scalar ($m_s$), and for $m_s \gtrsim 25$~\GeV\ they are almost independent of its mass and equal to 85\%, 5\%, and 8\%, respectively. While only these decay modes were considered in the previous ATLAS publication~\cite{EXOT-2017-05}, this paper also considers the decay into $t\overline{t}$ as it is the dominant decay mode when it becomes kinematically available.
 
The branching fraction for $\varPhi$ decaying into a pair of long-lived scalars is not constrained in these models. It is therefore interesting to focus on $\varPhi$ decays into LLPs where $\varPhi$ is a Higgs boson with a mass of 125~\GeV\ or another scalar with a different mass.
 
The relative masses and lifetimes generated, as well as details about the Monte Carlo (MC) event generation are described in Section~\ref{sec:datamc}. Only the gluon--gluon fusion production mode is considered, as it is the dominant production mode.

\section{Data and MC simulation}
\label{sec:datamc}
 
The analysis presented in this paper uses $\sqrt{s} = 13$~\TeV\ $pp$ collision data recorded by the ATLAS detector with stable LHC beams during the 2015--2018 data-taking periods~\cite{DAPR-2018-01}. Events are required to have been taken during stable beam conditions and when all the detector subsystems were operational. After data quality requirements~\cite{DAPR-2018-01}, the total integrated luminosity is 139~fb$^{-1}$.
 
An unbiased random trigger (the zero-bias trigger) that fires on the bunch crossing occurring one LHC revolution after a low-threshold calorimeter-based trigger was used to collect data for background estimation with negligible signal contamination. The zero-bias trigger runs throughout the ATLAS data-taking, so the events are acquired with the same beam conditions present in normal physics data-taking and can be used to study the expected background.
The zero-bias trigger was prescaled as a function of the instantaneous pileup to have a fixed output event rate throughout the data-taking period.
Therefore, to estimate the correct number of background events, the pileup profile from the zero-bias trigger was rescaled to match that from the data used to select candidate events.
 
Signal MC simulation samples were produced using the Hidden Abelian Higgs Model~\cite{NonSMCoupling2}, considering only the gluon--gluon fusion production mechanism. The masses, summarized in Table~\ref{tab:MCSamples}, were chosen to span the accessible parameter space. The mean proper lifetime in each sample was tuned to maximize the number of decays throughout the ATLAS detector volume and covers a range of 0.13--6~m, depending on the sample. All samples were generated at leading order using \MGNLO~\cite{MadGraph5} interfaced to the \PYTHIA[8]~\cite{PYTHIA8} parton shower model. The $\varPhi$ transverse momentum distribution for the samples is reweighted to match that obtained for the corresponding next-to-leading order (NLO) Higgs samples generated using the \MGNLO merging approach~\cite{Frederix:2012ps}. A set of tuned parameters called the A14 tune~\cite{ATL-PHYS-PUB-2014-021} was used together with the \NNPDF[3.1lo] parton distribution function (PDF) set~\cite{Ball:2012cx}. The \EVTGEN[1.2.0] program~\cite{EvtGent} was used to model $b$- and $c$-hadron decays. The generated events were processed through a full simulation of the ATLAS detector geometry and response~\cite{GEANT4b} using the \GEANT~\cite{GEANT4a} toolkit.
The simulation includes multiple $pp$ interactions per bunch crossing (pileup), as well as the effect on the detector response due to interactions from bunch crossings before or after the one containing the hard interaction.
Pileup was simulated with the soft strong-interaction processes of \PYTHIA[8] using the A3 tune~\cite{ATL-PHYS-PUB-2016-017} and the \NNPDF[2.3lo]~\cite{Ball:2012cx} PDF set.
Per-event weights were applied to the simulated events to match the distribution of the average number of interactions per bunch crossing measured in data.
 
A multijet background sample was generated using \PYTHIA[8.230]~\cite{Sjostrand:2014zea} with leading-order matrix elements for dijet production which were matched to the parton shower.
The \NNPDF[2.3lo] PDF set was used in the matrix element generation, the parton shower, and the simulation of the multiparton interactions. The A14 tune was used. Perturbative uncertainties were estimated through event weights~\cite{Mrenna:2016sih} that encompass variations of the scales at which the strong coupling constant is evaluated in the initial- and final-state showers, as well as the PDF uncertainty in the shower and the nonsingular part of the splitting functions.

\begin{table}[h!]
\centering
\caption{Mass parameters for the simulated scalar portal based on the Hidden Abelian Higgs Model~\cite{NonSMCoupling2}. A second set of samples with a different lifetime was generated for some scalar masses to validate the lifetime-extrapolation procedure.}
\begin{tabular}{ccl}
\toprule
$m_\Phi$ [\GeV]           & $m_s$ [\GeV]     & Proper lifetime [m] \\
\hline
\multirow{4}{*}{125}      & ~~5                &  0.127, 0.411  \\
& 16               &  0.580  \\
& 35               &  1.310, 2.630  \\
& 55               &  1.050, 5.320  \\ \cline{1-3}
\multirow{2}{*}{~~60}      & ~~5                &  0.217  \\
& 15               &  0.661 \\ \cline{1-3}
\multirow{1}{*}{200}     & 50               &  1.255  \\ \cline{1-3}
\multirow{1}{*}{400}     & 100~~              &  1.608 \\  \cline{1-3}
\multirow{3}{*}{600}     & 50               &  0.590 \\
& 150~~              &  1.840, 3.309  \\
& 275~~              &  4.288  \\  \cline{1-3}
\multirow{3}{*}{1000~~}    & 50               &  0.406  \\
& 275~~              &  2.399, 4.328  \\
& 475~~              &  6.039  \\
\bottomrule
\end{tabular}\label{tab:MCSamples}
\end{table}
 
Since the analysis is sensitive to a wide range of mean proper lifetimes, and the generation of many samples to cover a broad lifetime range would consume far too much CPU time, a toy MC method was adopted to extrapolate the expected number of events to the range of mean proper lifetimes between 0.01 and 1000~m. For each LLP in the MC sample, a random decay time, sampled from an exponential distribution of a chosen proper lifetime, was generated. The physical decay distance in the detector was then calculated for each simulated LLP using its four-momentum. The overall probability of the event to satisfy the signal selection criteria, parameterized as a function of the LLP decay position and boost, is then evaluated from the LLP trigger and vertex efficiencies described in Sections~\ref{sec:RoICluTrig} and \ref{sec:2MSVxEff}, respectively. In order to validate the lifetime-extrapolation procedure described above, a second set of samples with a different lifetime was generated for some scalar masses.

\section{Trigger selection and event reconstruction}
\label{sec:trigvtxreco}
 
Hadronic LLP decays in the MS typically produce narrow, high-multiplicity hadronic showers. The track multiplicity and shower width vary with the mass and boost of the decaying LLP and the final states to which the LLP decays. Dedicated trigger~\cite{MuonRoITrigger} and vertex~\cite{PERF-2013-01} algorithms were developed to select and reconstruct displaced decays in the MS. Due to the amount of material in the calorimeter, only decays occurring in or after the last sampling layer of the hadronic calorimeter produce enough hits in the MS to allow reconstruction of the DV.

\subsection{Muon RoI Cluster trigger}\label{sec:RoICluTrig}
 
The Muon RoI Cluster trigger is a signature-driven trigger that selects candidate events for decays of LLPs in the MS~\cite{MuonRoITrigger}. Events must contain at L1 two muon RoIs with a \pT higher than 10~\GeV, and in the HLT a cluster of three (four) muon RoIs within a $\Delta R=0.4$ cone centered on the L1 object in the barrel (endcaps). The isolation criteria for jets and tracks discussed in Ref.~\cite{MuonRoITrigger}, used to reduce background DVs from punch-through jets, are not considered in the analysis presented in this paper.
The trigger selects both isolated, signal-like events and nonisolated, \mbox{background-like} events used for data-driven background estimations.

\begin{figure}[htbp]
\begin{subfigure}[b]{0.49\textwidth}
\includegraphics[width=\textwidth]{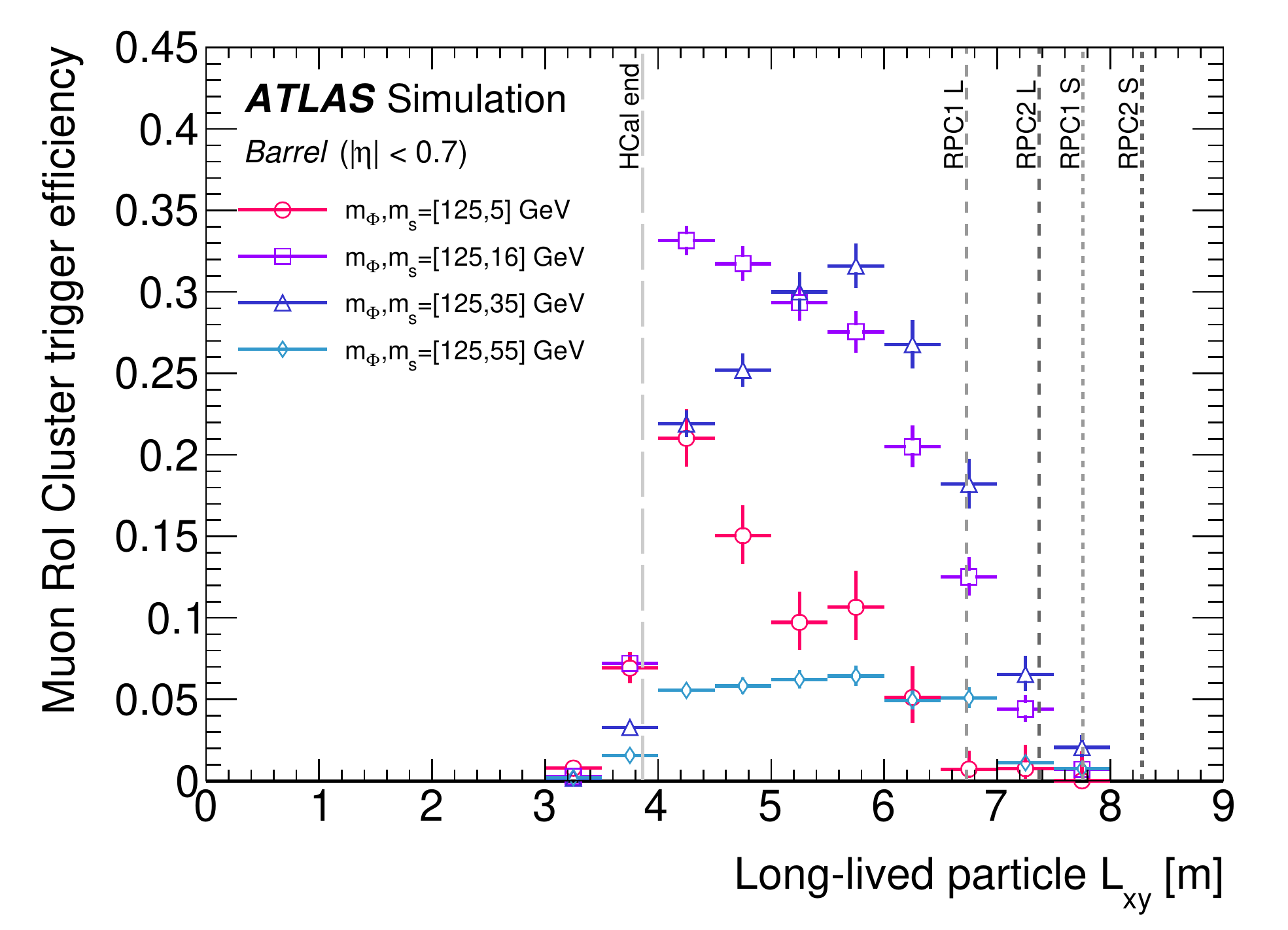}
\caption{}\label{fig:MuonClusterTriggerEffBarrel}
\end{subfigure}
\begin{subfigure}[b]{0.49\textwidth}
\includegraphics[width=\textwidth]{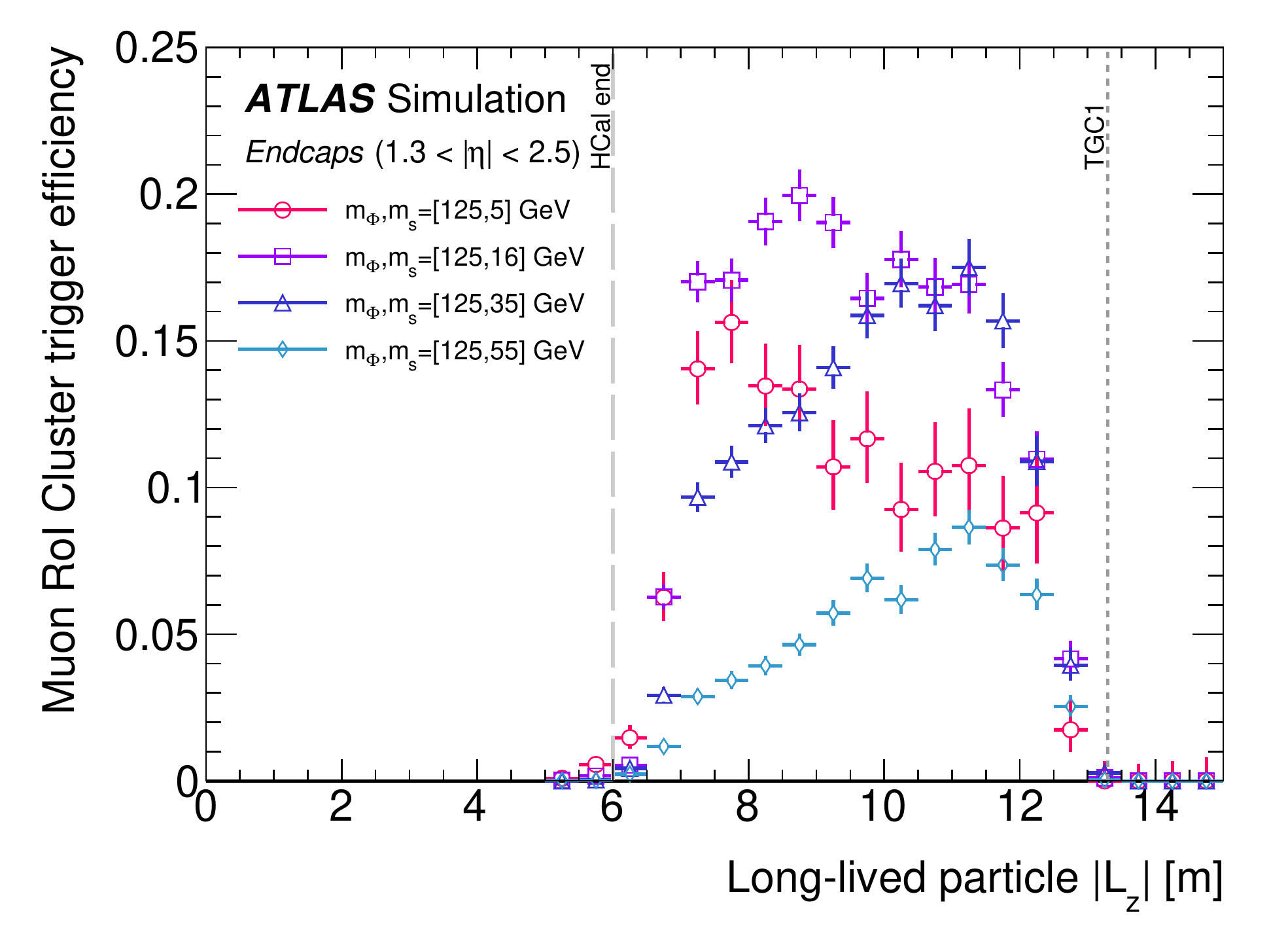}
\caption{}\label{fig:MuonClusterTriggerEffEndcaps}
\end{subfigure}
\caption{Efficiency for the Muon RoI Cluster trigger as a function of the decay position of the LLP for some scalar portal samples in the (a) MS barrel and (b) MS endcaps for events passing the data quality requirements and having a reconstructed primary vertex. These efficiency distributions are based solely on MC simulation, without any corrections applied for the mismodeling described in Section~\ref{sec:RoICluTrig}. The vertical lines show the relevant detector boundaries, where ``HCal end'' is the outer limit of the hadronic calorimeter, ``RPC 1/2'' represent the first/second stations of RPC chambers, ``TGC 1'' represents the first stations of TGC chambers and ``L/S'' indicate whether they are in the Large or Small sectors.}
\label{fig:MuonClusterTriggerEff}
\end{figure}

The trigger efficiency, defined as the fraction of LLP decays in the MS fiducial volume that are selected by the trigger as a function of the LLP decay position, is shown in Figures~\ref{fig:MuonClusterTriggerEffBarrel} and~\ref{fig:MuonClusterTriggerEffEndcaps} for four MC simulated benchmark samples with LLP decays in the MS barrel and endcap regions, respectively. The efficiency is parameterized as a function of the transverse decay radius ($L_{xy}$) in the barrel and the longitudinal decay position ($|L_z|$) in the endcaps. The events are required to pass the data quality requirements~\cite{DAPR-2018-01} and have a reconstructed primary vertex as described in Section~\ref{sec:PVJetReco}. The trigger is efficient for hadronic decays of LLPs that occur anywhere from the outer regions of the HCal to the middle stations of the MS. These efficiencies are obtained from a subset of simulated signal events with only a single LLP decay in the muon spectrometer in order to ensure that the result of the trigger is due to a single burst of MS activity. The uncertainties shown are statistical only. The relative differences between the efficiencies of the benchmark samples are a result of the different masses of the LLPs, which in turn affect their momenta and consequently the opening angles of the decay products. The trigger efficiency increases as the LLPs decay closer to the end of the hadronic calorimeter, located at $r \approx 4$~m for the barrel and $|z| \approx 6$~m for the endcaps. The efficiency decreases rapidly as the decay occurs closer to the middle stations of the muon spectrometer (barrel: $r\approx 7~$m; endcaps: $|z| \approx 13$~m).

To quantify the mismodeling of the L1 muon trigger efficiency in MC simulation, the distributions of the number of muon-RoI clusters within a cone of $\Delta R = 0.4$ around the axis of a punch-through jet in multijet MC and data events are compared.  High-energy jets in data were selected using jet triggers with a \pT threshold of 400~\GeV\ or 420~\GeV, depending on the data-taking period.
MC simulation shows a rate of muon-RoI clusters that is 24\% (20\%) lower than in data in the barrel (endcaps). This mismodeling does not depend on the $\eta$, $\phi$ or \pT of the jet, and it is used as a systematic uncertainty on the signal trigger efficiency.

\subsection{Reconstruction of MS displaced vertices}\label{sec:vxReco}
 
A dedicated algorithm~\cite{PERF-2013-01}, capable of reconstructing low-momentum tracks in a busy environment, is used to reconstruct the MS DV. This algorithm was also employed by previous searches for displaced decays in the MS~\cite{HVAtlas,HVAtlas8TeV,EXOT-2017-05}. The algorithm takes advantage of the spatial separation between the two multilayers inside a single MDT chamber. Single-multilayer straight-line segments that contain three or more MDT hits are reconstructed using a $\chi^2$ fit. Segments from multilayer 1 are then matched with those from multilayer 2. The matched pair of a multilayer-1 segment with a multilayer-2 segment and their corresponding track parameters is called a \emph{tracklet}. Tracklets that are spatially clustered within $|\Delta\eta|<0.7$ and $|\Delta\phi|<\pi/3$ in the detector are extrapolated backwards to reconstruct the $(\eta, \phi)$ position of the MS DV with a $\chi^2$ fit. At least three (four) tracklets are required to be used in the fit in the barrel (endcaps).
Detectable vertices arise from decays that occur in the region between the outer edge of the HCal and the middle stations of muon chambers. Because their detector technology is different (no spatially separated multilayers), the CSC chambers are not used for the MS DV reconstruction.

\subsection{Reconstruction of the primary vertex and prompt hadronic jets}\label{sec:PVJetReco}
 
Events are required to have a primary vertex (PV) with at least two tracks with $\pT > 500$~\MeV. If more than one PV candidate is reconstructed, the candidate with the largest sum of the squares of the transverse momenta of all tracks associated with the vertex is selected.
 
Hadronic jets are constructed by using FastJet~\cite{Fastjet} to apply the \antikt jet algorithm \cite{Cacciari:2008gp} with a radius parameter $R=0.4$. A collection of three-dimensional topological clusters of neighboring energy deposits in the calorimeter cells containing a significant energy above a noise threshold~\cite{PERF-2014-07} provide input to the \antikt algorithm. After reconstruction, jets are calibrated using the procedure outlined in Refs.~\cite{PERF-2012-01, PERF-2011-03, PERF-2014-03, ATLAS-CONF-2015-002, FTAG-2018-01, ATL-PHYS-PUB-2015-015, PERF-2016-04}.

\section{Event selection}
\label{sec:eventselection}
 
All events considered in this search must pass an event-level selection to distinguish signal from background.

\subsection{Baseline event selection}
\label{sec:baseselection}
 
A common baseline selection is applied to the events considered in this search and summarized in Table~\ref{tab:baselineSelection}.
 
\begin{table}[!h]
\centering
\caption{Summary of the baseline selection criteria applied to data and simulated events. The variables $n_{\textrm{MDT}}$/$n_{\textrm{RPC}}$/$n_{\textrm{TGC}}$ are the numbers of MDT/RPC/TGC hits in the vertex cone, as described in the text, and $\eta_{\textrm{vx}}$ is the pseudorapidity of the MS DV with respect to the IP.}
\begin{tabular}{p{6 cm}p{6 cm}}
\toprule
\multicolumn{2}{l}{Event passes data quality requirements and Muon RoI Cluster trigger} \\
\midrule
\multicolumn{2}{l}{Event has a PV with at least two tracks with $\pT > 500$~\MeV} \\
\midrule
\multicolumn{2}{l}{Event has at least one MS DV} \\
\midrule
\multicolumn{2}{l}{MS DV is matched to the triggering muon-RoI cluster ($\Delta R(\text{DV}, \text{RoI cluster})<0.4$)} \\
\multicolumn{2}{l}{In the case of two muon-RoI clusters, the second vertex must be} \\
\multicolumn{2}{l}{matched to the second cluster} \\
\midrule
\multicolumn{2}{l}{300 $\le n_{\textrm{MDT}} <$ 3000} \\
\bottomrule
\end{tabular}
 
\vspace{0.1 cm}
 
\begin{tabular}{p{5 cm}p{7.5 cm}}
\toprule
\centering \emph{Barrel} & \centering \emph{Endcaps} \tabularnewline
\midrule
MS DV with $|\eta_{\textrm{vx}}| < 0.7$  & MS DV with $1.3 < |\eta_{\textrm{vx}}| < 2.5$ \\
MS DV with $3\,\text{m} < L_{xy} < 8\,\text{m}$ & MS DV with $L_{xy} < 10\,\text{m}$ and $5\,\text{m} < |L_z| < 15\,\text{m}$ \\
$n_{\textrm{RPC}} \geq 250$ & $n_{\textrm{TGC}} \geq 250$ \\
\bottomrule
\end{tabular}
\label{tab:baselineSelection}
\end{table}
 
Events with at least one MS DV are required to pass the data quality requirements~\cite{DAPR-2018-01} and the Muon RoI Cluster trigger, and contain a PV. The PV selection has very little impact on the signal efficiency but it helps to reject background events; in simulation, the selected PV corresponds to the signal interaction in about 95\%--99\% of the cases, depending on the sample. Even though the LLPs are invisible in the ID, the scalar boson is produced with an average \pT of 20~\GeV, and a number of tracks are left by recoiling particles produced in the same $pp$ interaction.
 
An MS DV that arises from a displaced decay typically has many more hits than an MS DV from background. To take advantage of this difference, a minimum number of MDT ($n_{\textrm{MDT}}$) and RPC/TGC ($n_{\textrm{RPC}}$/$n_{\textrm{TGC}}$) hits is required. The $n_{\textrm{MDT}}$ hits are counted in the MDT chambers that have their center within $\Delta \phi = 0.6$ and $\Delta \eta = 0.6$ of the DV $(\eta,\phi)$ direction. The $n_{\textrm{RPC}}$ and $n_{\textrm{TGC}}$ hits are the sum of hits that are within $\Delta R = 0.6$ of the DV position. A requirement on the maximum number of MDT hits is also applied to remove background events caused by coherent noise bursts in the MDT chambers. This selection has a negligible impact on the signal events.
 
A displaced decay that occurs in the transition region between the MS barrel and endcaps results in hits in both regions. Vertex reconstruction is performed separately in the barrel and endcaps, and only the barrel (endcap) hits are used in the barrel (endcap) vertex reconstruction algorithm. Therefore, MS  vertices reconstructed from either of the two algorithms have fewer hits, as they are reconstructed from a subset of the full set of hits. This results in a decrease in the reconstruction efficiency, and occasionally two vertices being reconstructed from a single LLP decay. Therefore, the MS DVs with pseudorapidity $|\eta_{\textrm{vx}}|$ between 0.8 and 1.3 are not considered in the analysis. This has a negligible effect on the total signal efficiency, because the average MS DV reconstruction efficiency in this region is less than 2\%. Moreover, in the transition region between the barrel and the endcap hadronic calorimeters, $0.7 < |\eta_{\textrm{vx}}| < 1.2$, the probability of having a jet that does not fulfill the minimal selection criteria to be considered for isolation and also punches through into the MS is much higher than in other regions of the detector. This region overlaps the already excluded MS transition region, except for  $0.7 < |\eta_{\textrm{vx}}| < 0.8$.
Therefore, vertices reconstructed with pseudorapidity in the range $0.7 < |\eta_{\textrm{vx}}| < 0.8$ are also not considered.

The number of events passing the baseline selection is 1,385,587 and 5,138,794 in the barrel and endcaps, respectively. After these requirements, the main background contribution is from punch-through jets.

\subsection{Signal displaced vertex selection}\label{sec:2MSVxSelection}

To reject background vertices created by punch-through jets, a set of \emph{vertex isolation criteria} was established. These criteria are based on the angular distance, $\Delta R$, between the direction of the tracks or jets and the vertex axis, defined as the line from the IP to the DV.
No jets or tracks should be present in a $\Delta R$ cone around the MS DV axis.
For isolation from tracks, two criteria are used. One is for isolation from tracks with $\pT>5$~\GeV\, (high-\pT tracks). The other takes the vector \pT sum of all tracks associated with the primary vertex (PV) that have $500~\MeV<\pT<5$~\GeV\ (low-\pT tracks) and are in a cone of $\Delta R = 0.2$ around the MS DV axis . The use of two different isolation criteria stems from the fact that some jets have most of their energy in a few hadrons, while others can consist of multiple low-\pT tracks. The isolation criteria are summarized in Table~\ref{tab:vertexIso}. An MS DV that satisfies these criteria is considered in the analysis.
 
\begin{table}[h!]
\centering
\caption{Summary of the vertex isolation criteria used to select signal DVs in the barrel and endcap regions.}
\begin{tabular}{lcc}
\toprule
\multicolumn{1}{c}{Isolation requirements} & Barrel & Endcaps \\
\midrule
Isolation from high-\pT tracks ($\pT > 5$~\GeV) & $\Delta R > 0.3$ & $\Delta R >$ 0.6 \\
Isolation from low-\pT tracks $(\Sigma \pT(\Delta R<0.2))$ & $\Sigma \pT < 10$~\GeV\ & $\Sigma \pT < 10$~\GeV\ \\
Isolation from jets & $\Delta R > 0.3$ & $\Delta R > 0.6$  \\
\bottomrule
\end{tabular}
\label{tab:vertexIso}
\end{table}
 
\begin{figure}[htbp!]
\begin{subfigure}[b]{0.49\textwidth}
\includegraphics[width=\textwidth]{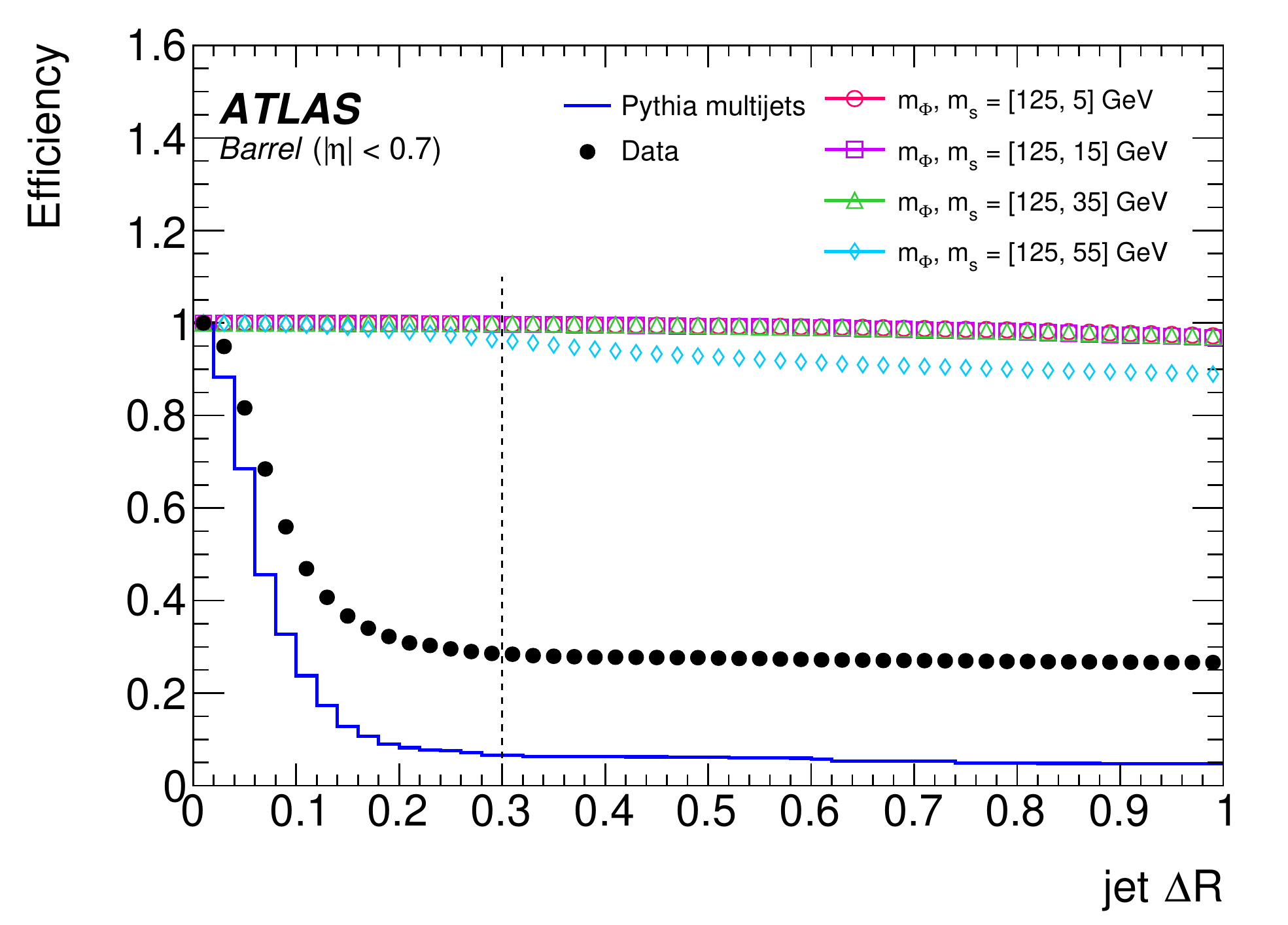}
\caption{}\label{fig:GVCjetdRBarrel}
\end{subfigure}
\begin{subfigure}[b]{0.49\textwidth}
\includegraphics[width=\textwidth]{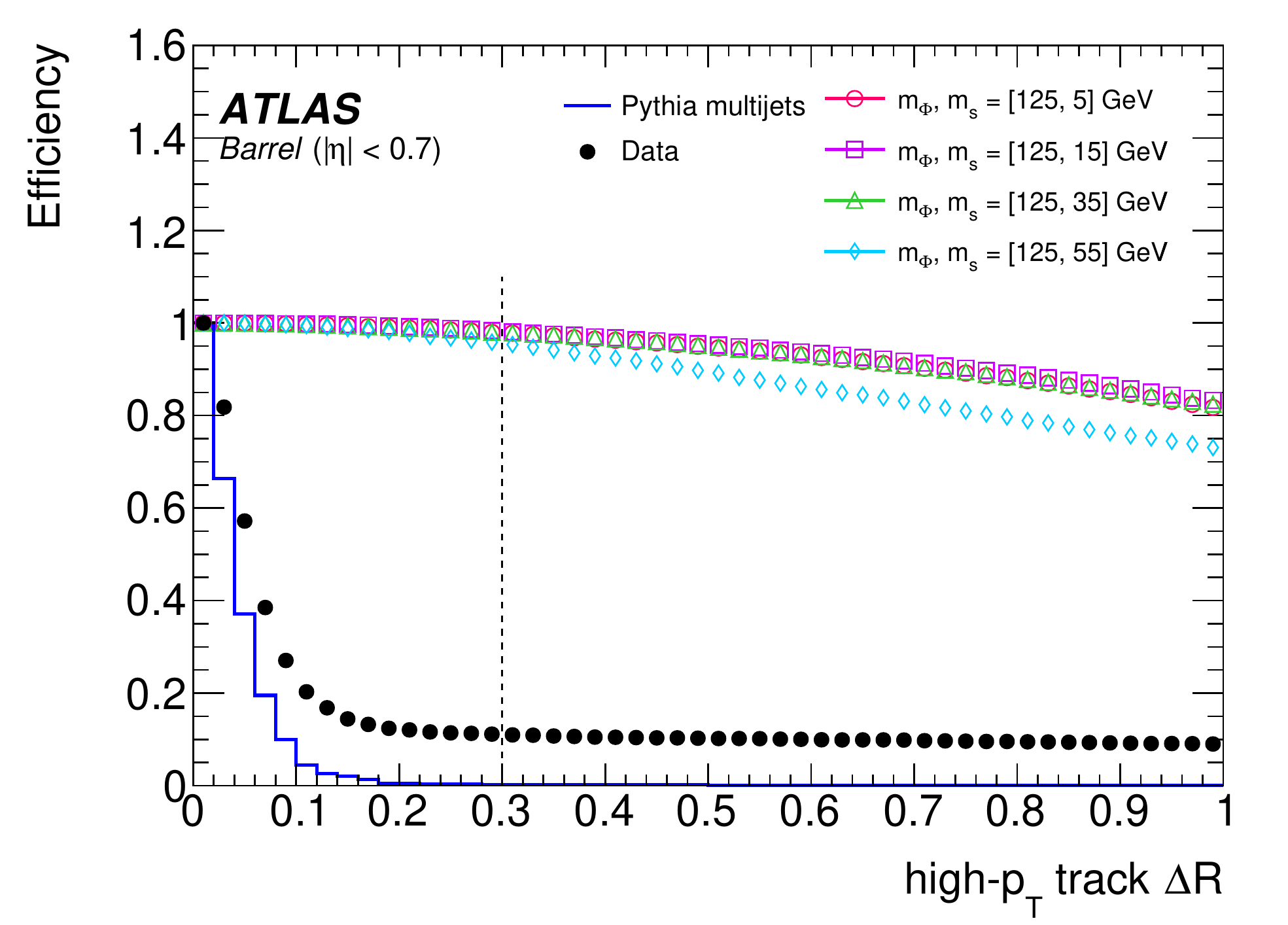}
\caption{}\label{fig:GVCtrackdRBarrel}
\end{subfigure}
\begin{subfigure}[b]{0.49\textwidth}
\includegraphics[width=\textwidth]{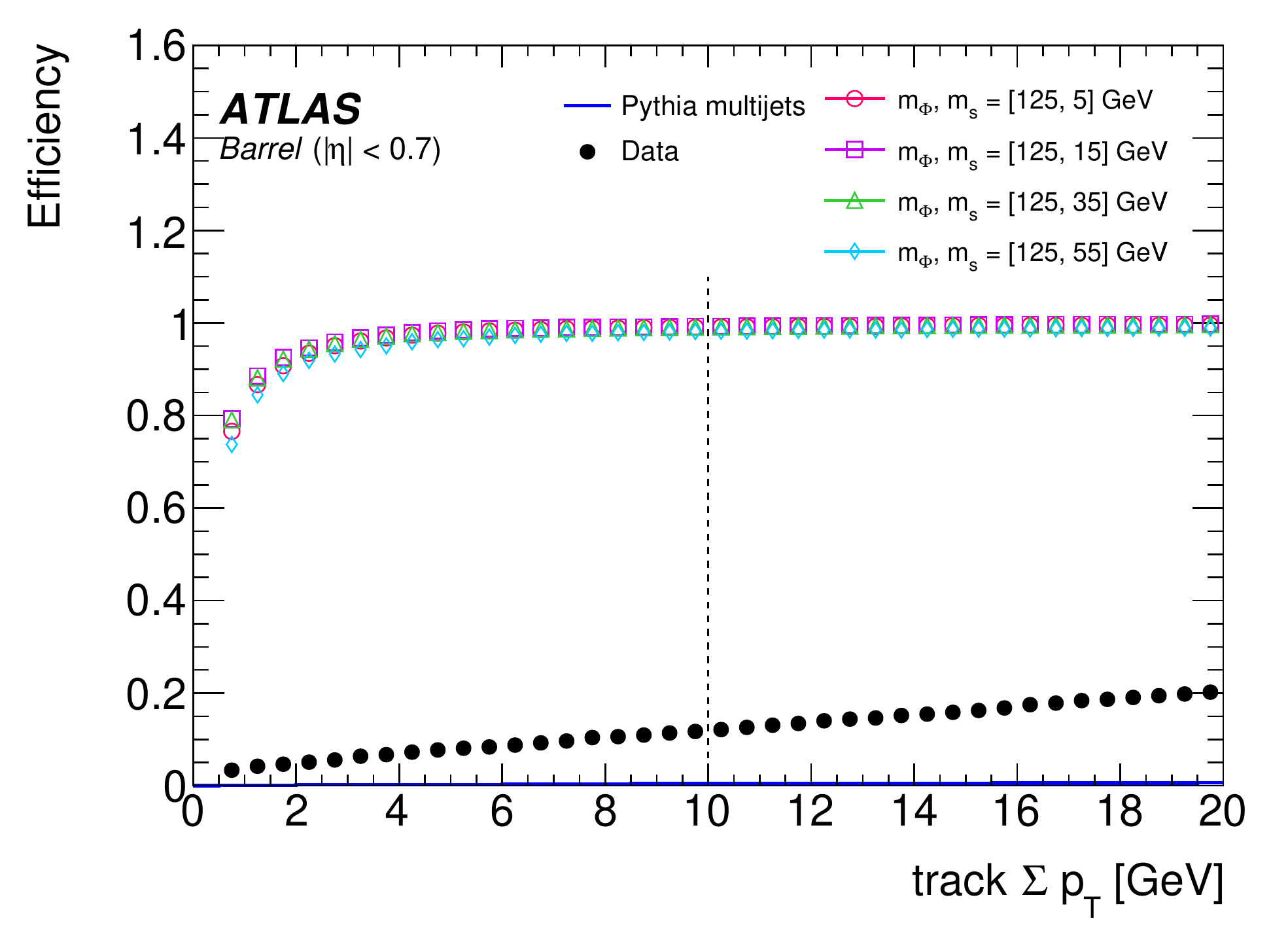}
\caption{}\label{fig:GVCsumtrackpTBarrel}
\end{subfigure}
\begin{subfigure}[b]{0.49\textwidth}
\includegraphics[width=\textwidth]{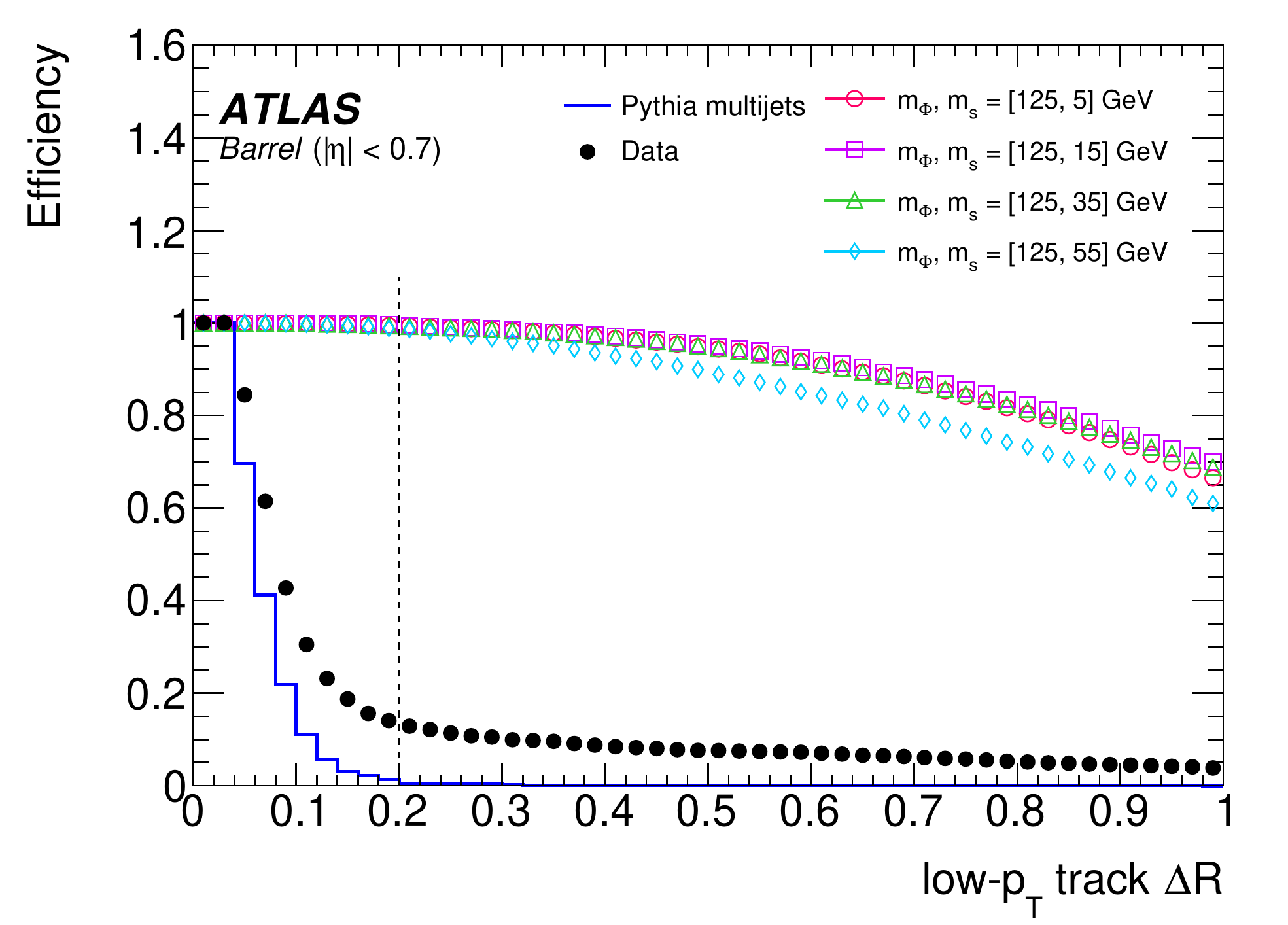}
\caption{}\label{fig:GVCnMDTBarrelUpper}
\end{subfigure}
\caption{Cumulative efficiency of vertices in the barrel where the vertex is required to be isolated from (a) jets and (b) high-\pT tracks, as a function of the selected $\Delta R$, when the sum of low-\pT tracks in a $\Delta R = 0.2$ cone around the vertex direction is required to be less than a specified cut in \pT (c), and when the sum of low-\pT tracks in a $\Delta R$ cone around the vertex is required to be less than $10~\GeV$, as a function of the $\Delta R$ value (d). The vertical lines show the selected $\Delta R$ value that is used in analysis. The differences between Pythia multijet and data distributions are attributable to the noncollision background, which is not present in MC simulations.}
\label{fig:GVCPlotsBarrel}
\vspace{-0.09 cm}
\end{figure}

The isolation criteria were optimized for the MC benchmark samples by comparing simulated signal events with simulated multijet events.
Figure~\ref{fig:GVCPlotsBarrel} shows the cumulative vertex efficiency as a function of the isolation requirements in the barrel for data events and simulated multijet and signal events.
 
All jets considered for DV isolation must satisfy $\pT>30$~\GeV\ and $\LogCalRatio<0.5$, where $E_{\textrm{HAD}}$ is the jet energy deposited in the HCal, and $E_{\textrm{EM}}$ the energy deposited in the ECal. The latter criterion is commonly used to identify the decay of a neutral LLP in the hadronic calorimeter~\cite{EXOT-2017-25} that results in little or no energy deposited in the electromagnetic calorimeter and consequently in an anomalously large value of the hadronic to electromagnetic energy ratio. It ensures that vertices originating from LLPs that decay near the outer edge of the hadronic calorimeter and also have significant MS activity are not rejected by the isolation requirement. In addition, in order to reduce the probability that pileup jets prevent a signal vertex from meeting the isolation criteria, jets with $20 < \ET < 60$~\GeV\ must be matched to the PV by using a jet vertex tagger (JVT) discriminant~\cite{PERF-2014-03}. Standard jet-quality criteria~\cite{ATLAS-CONF-2015-029} are not applied because jets that do not fulfill these requirements can also produce a background MS DV, and therefore need to be considered when computing the isolation.
 
At least two isolated MS DVs must be present in the event. One MS DV must be matched to the trigger-level muon-RoI cluster by satisfying $\Delta R(\textrm{cluster, vertex}) < 0.4$. If there are two distinct clusters, each MS DV must be matched to a different cluster. To ensure that the two MS DVs and/or two muon-RoI clusters do not come from the same background activity, the two vertices are required to be separated by at least $\Delta R = 1.0$, which has minimal impact (${\lesssim}3$\%) on the overall signal acceptance.  After the vertex isolation criteria are applied, most vertices from punch-through jets are eliminated, and the main background contribution is from noncollision backgrounds.

\subsection{MS displaced vertex reconstruction efficiency}\label{sec:2MSVxEff}
 
\begin{figure}[htbp]
\begin{subfigure}[b]{0.49\textwidth}
\includegraphics[width=\textwidth]{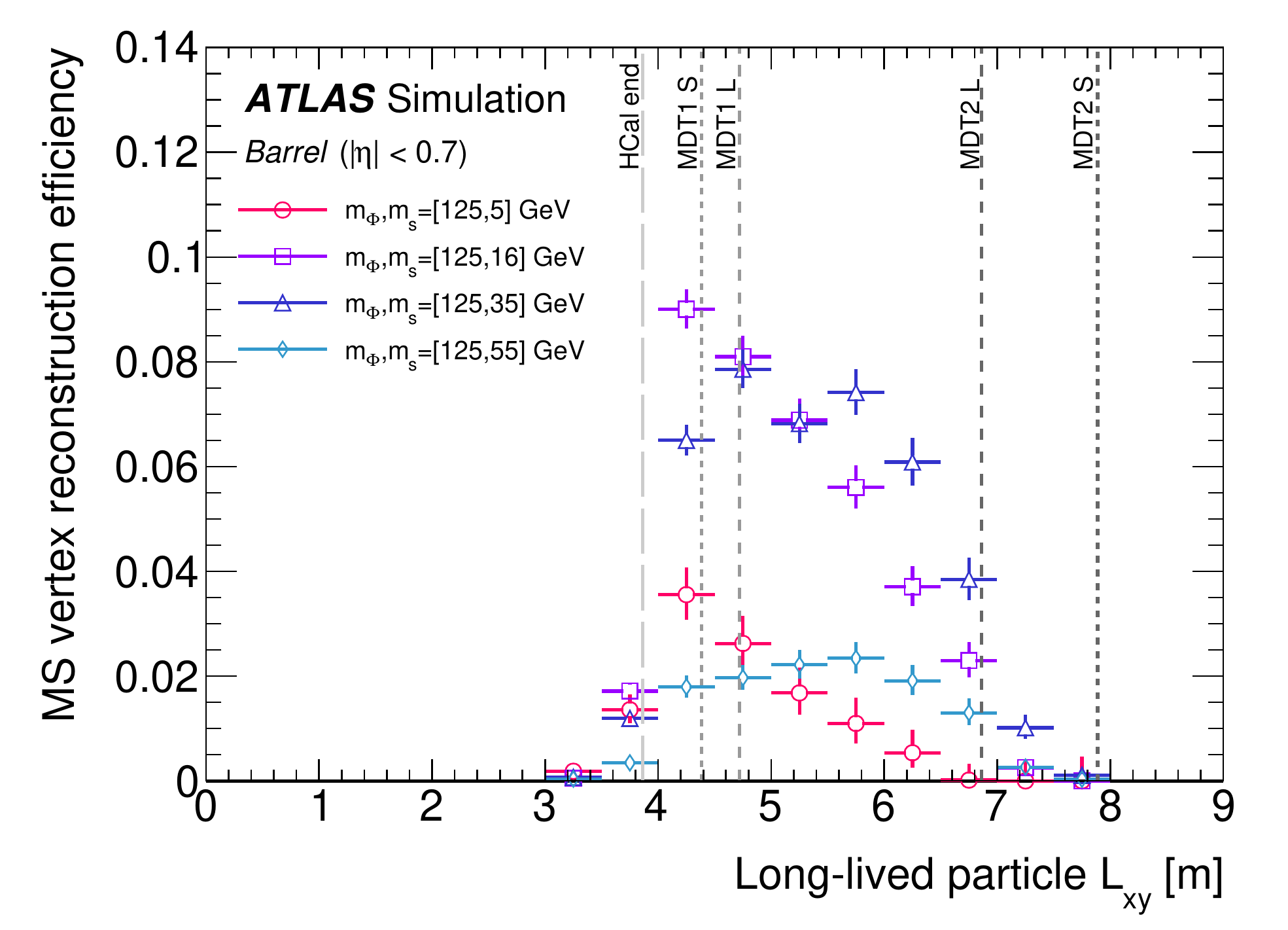}
\caption{}
\label{fig:MSTriggerVertexEffBarrel}
\end{subfigure}
\begin{subfigure}[b]{0.49\textwidth}
\includegraphics[width=\textwidth]{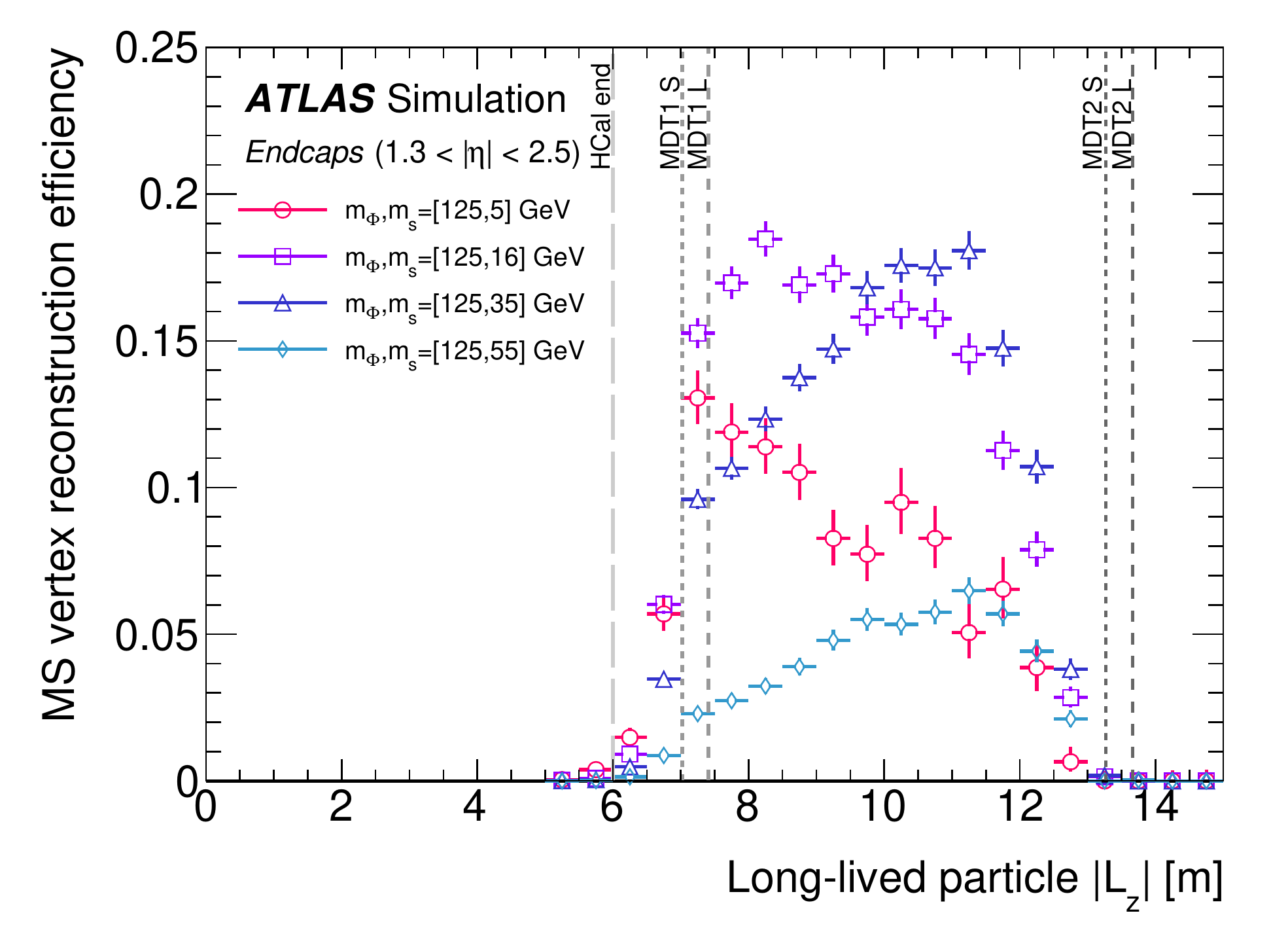}
\caption{}
\label{fig:MSTriggerVertexEffEndcap}
\end{subfigure}
\caption{Efficiency to reconstruct an MS DV for scalar portal samples with $m_\varPhi=125$~\GeV\ for vertices that pass the baseline event selection and satisfy the vertex isolation criteria. (a): Barrel MS DV reconstruction efficiency as a function of the transverse decay position of the LLP. (b): Endcap MS DV reconstruction efficiency as a function of the longitudinal decay position of the LLP relative to the center of the detector. The efficiency distributions are corrected for the mismodeling described in Section~\ref{sec:2MSVxEff}. The vertical lines show the relevant detector boundaries, where ``HCal end'' is the outer limit of the hadronic calorimeter, ``MDT 1/2'' represent the first/second stations of MDT chambers and ``L/S'' indicate whether they are in Large or Small sectors.}
\label{fig:MSTriggerVertexEffs}
\end{figure}

The efficiency for vertex reconstruction~\cite{PERF-2013-01} is defined as the fraction of simulated LLP decays in the MS fiducial volume that correspond to a reconstructed vertex passing the baseline event selection reported in Table~\ref{tab:baselineSelection} and satisfying the vertex isolation criteria summarized in Table~\ref{tab:vertexIso}. A reconstructed vertex is considered matched to a displaced decay if the vertex is within $\Delta R = 0.4$  of the simulated decay position. The MS DV efficiency is parameterized as a function of the $L_{xy}$ and $|L_z|$ LLP decay position in the barrel and endcaps, respectively. Figure~\ref{fig:MSTriggerVertexEffBarrel} shows the efficiency for reconstructing a vertex in the MS barrel for a selection of benchmark samples. Figure~\ref{fig:MSTriggerVertexEffEndcap} shows the efficiency for reconstructing a vertex in the MS endcaps.
 
For the MC samples considered in this paper, the MS barrel vertex reconstruction efficiency is $\mathcal{O}(2-15)$\% near the outer edge of the hadronic calorimeter ($r\approx4$\,m) and decreases substantially as the decay occurs closer to the middle stations ($r\approx7$\,m). The decrease occurs because the charged hadrons and photons are not spatially separated and overlap when they traverse the middle stations. This results in a reduction of the efficiencies for tracklet reconstruction and, consequently, vertex reconstruction.
The efficiencies are also shaped by the mass and boost of the LLP.
The efficiency for reconstructing vertices is higher in the MS endcaps due to a more efficient selection and vertex reconstruction, and it reaches 40\% for higher-mass benchmark models. Since the magnetic field in the region in which endcap tracklets are reconstructed is very weak, the vertex reconstruction does not have the curvature constraints on the tracklets from the charge and momentum measurements that are  present in the barrel. Therefore, in the endcaps, the vertex reconstruction algorithm uses straight-line fits so that low-momentum tracks are not rejected, while in the barrel the curvature plus combinatorics provide better rejection of misreconstructed tracks. Consequently, the vertex reconstruction in the endcaps is more efficient for signal, but also less robust in rejecting background events. More details are provided in Ref.~\cite{PERF-2013-01}.
 
Potential MC simulation inaccuracies in the DV reconstruction are estimated by comparing the distribution of the number of tracklets in a punch-through jet $\Delta R$ cone of $0.4$ in data and MC events using a strategy similar to the one used for the trigger, which was described in Section~\ref{sec:RoICluTrig}. MC simulation shows a rate of tracklets that is 20\% (15\%) higher than in data in the barrel (endcaps). This mismodeling does not depend on the $\eta$, $\phi$ or \pT of the jet. Its impact on the DV reconstruction efficiency is estimated by randomly dropping tracklets used to reconstruct the vertex in accordance with the measured mismodeling factor in the barrel and endcaps and counting the number of reconstructed vertices. The efficiency variation between the nominal reconstruction and the one considering the mismodeling factors, averaged over the MC benchmark samples, is 27\% in the barrel, and 9\% in the endcaps. The corrected efficiencies are used to compute the expected number of signal events. The systematic uncertainty associated to this correction is discussed in Section~\ref{sec:syst}.

Over the range of the extrapolated lifetimes, the maximum acceptance times efficiency for vertices passing the signal selection ranges from 0.005\% (for the MC sample with $m_\Phi = 60$~\GeV, $m_s = 5$~\GeV\ at a proper lifetime of 0.5 m) to 0.3\% (for the MC sample with $m_\Phi = 600~\GeV$, $m_s = 50$~\GeV\ at a proper lifetime of 1.1 m).
 
\section{Background estimation}
\label{sec:Bkg2MSVx}
 
The two-MS-vertex strategy is designed to be sensitive to models where two LLPs are produced and decay hadronically between the outer region of the HCal and the middle stations of the MS. Requiring two MS DVs significantly reduces the background. In addition, background from punch-through jets is further reduced using the isolation criteria described in Section~\ref{sec:2MSVxSelection}.
 
Residual background can arise from collision or noncollision processes and cannot be simulated accurately. Thus, a data-driven method is used to estimate the expected background.
This has the advantage of avoiding the systematic uncertainties related to simulation accuracy present when using MC-based background estimates.

As described in Section~\ref{sec:baseselection}, the signal selection requires that one vertex is always matched to a RoI cluster, and only in the case of two RoI clusters the second vertex is also required to be matched. Therefore, we can naturally factorizes the background estimation into three terms:
 
\begin{equation}
N_\textrm{2Vx} = N^\textrm{1cl} \cdot P^\textrm{Vx}_\textrm{noMStrig} +
N^\textrm{2cl}_\textrm{1UMBcl} \cdot P^\textrm{Vx}_\textrm{Bcl} +
N^\textrm{2cl}_\textrm{1UMEcl} \cdot P^\textrm{Vx}_\textrm{Ecl} \,\,\,\,\, .
\label{eq:1}
\end{equation}
 
The first term in Eq.~(\ref{eq:1}) gives the number of background events that contain one vertex-cluster pair anywhere in the detector fiducial region and one other isolated MS vertex anywhere in the same fiducial region. The second and third terms estimate the number of background events that contain one vertex-cluster pair anywhere in the detector fiducial region and an additional vertex-cluster pair in the barrel or endcap region.
The background from events with one muon-RoI cluster (first term in Eq.~(\ref{eq:1})) is estimated by multiplying the number of events with one muon-RoI cluster ($N^\textrm{1cl}$) by the probability of finding an isolated MS DV vertex in events not selected by the Muon RoI Cluster trigger ($P^\textrm{Vx}_\textrm{noMStrig}$). The probability $P^\textrm{Vx}_\textrm{noMStrig}$ is determined using data collected with the zero-bias trigger and it is found by dividing the number of events with an isolated MS DV ($N^\textrm{1\textrm{Vx}}$) by the total number of zero-bias events that satisfy standard event-quality requirements ($N_\textrm{Events}$). The background from events with two muon-RoI clusters (second and third terms in Eq.~(\ref{eq:1})) is estimated by multiplying the number of events with two muon-RoI clusters where one of the two clusters is not matched to an isolated vertex in the barrel ($N^\textrm{2cl}_\textrm{1UMBcl}$) or endcaps ($N^\textrm{2cl}_\textrm{1UMEcl}$) by the probability of finding an MS DV vertex given a muon-RoI cluster in the barrel or endcaps respectively ($P^\textrm{Vx}_\textrm{Bcl}$ and $P^\textrm{Vx}_\textrm{Ecl}$). The probabilities $P^\textrm{Vx}_\textrm{Bcl}$ and $P^\textrm{Vx}_\textrm{Ecl}$ are found by dividing the number of events with an isolated MS DV matched to a muon-RoI cluster ($N^{\textrm{1BclVx}}$ and $N^{\textrm{1EclVx}}$) by the number of events with a muon-RoI cluster ($N^{\textrm{1Bcl}}$ and $N^{\textrm{1Ecl}}$).

All the numbers of events and probabilities used in Eq.~(\ref{eq:1}) are reported in Table~\ref{tab:backgroundNums}.
 
\begin{table}[h!]
\centering
\caption{Numbers of events in the 13~\TeV\ dataset passing the zero-bias and Muon RoI Cluster triggers.
}
\begin{tabular}{llc}
\toprule
Stream & Quantity & Value \\
\midrule
\multirow{3}{*}{Zero-bias} & $N_{\textrm{Events}}$               &  $115,709,000 \pm 11,000$  \\
& $N^{\textrm{1Vx}}$                    &  $53 \pm 7$ \\
& $\qquad \rightarrow  P^\textrm{Vx}_\textrm{noMStrig}$ & $(4.58 \pm 0.63) \times 10^{-7}$ \\
\midrule
\multirow{9}{*}{Muon RoI Cluster} & $N^\textrm{1cl}$     & $674,800 \pm 800$ \\
& $N^\textrm{2cl}_\textrm{1UMBcl}$ & $3^{+2.43}_{-1.17}$  \\
& $N^\textrm{2cl}_\textrm{1UMEcl}$ & $0^{+1.35}_{-0.00}$  \\
& $N^{\textrm{1Bcl}}$     & $38,509,000 \pm 6000$  \\
& $N^{\textrm{1BclVx}}$ & $124,650 \pm 350$  \\
& $\qquad \rightarrow P^\textrm{Vx}_\textrm{Bcl} $ & $(3.24 \pm 0.01) \times 10^{-3}$  \\
& $N^{\textrm{1Ecl}}$     & $15,599,000 \pm 4000$  \\
& $N^{\textrm{1EclVx}}$ & $550,100 \pm 700$  \\
& $\qquad \rightarrow P^\textrm{Vx}_\textrm{Ecl} $ & $(3.53 \pm 0.01) \times 10^{-2}$  \\
\bottomrule
\end{tabular}
\label{tab:backgroundNums}
\end{table}
 
All events used for the background estimation are orthogonal to the signal region that requires two isolated DVs. Therefore, events that pass the Muon RoI Cluster trigger and have only one isolated MS DV are assumed to be fake vertices not due to displaced decays. Potential signal contamination would result in an overestimation of the background rates, but its contribution has been estimated to be smaller than 0.02\% for injected signal with rates of the order of the observed 95\% CL upper limits.

The background estimation strategy was validated using an orthogonal signal-free data sample obtained by inverting the isolation criteria used to define the signal region. In this validation region, the expected number of background events was calculated to be $0.99 \pm 0.20$ and $1.56 \pm 0.27$ for one-cluster and two-cluster events, respectively, where the uncertainty is purely statistical. One event, which contained two clusters, was observed. Therefore, no systematic error is assigned to the background estimation.

The number of expected background events with two MS DVs is $0.32 \pm 0.05$, where the uncertainty is statistical only.

\section{Systematic uncertainties}
\label{sec:syst}

The signal efficiency systematic uncertainties are dominated by the modeling of the signal physics processes, pileup and detector response and the extrapolation of the expected number of signal events as a function of the LLP proper lifetime.
 
One of the sources of systematic uncertainty associated with the Muon RoI Cluster trigger is the modeling of the minimum-bias interactions used to simulate pileup; another stems from the systematic uncertainty due to the PDF used to generate signal MC events. These were estimated by varying the pileup and PDF weights in accord with the respective $\pm 1\sigma$ systematic uncertainties and evaluating the resultant change in the trigger efficiency. For the latter, uncertainties in the nominal PDF set were evaluated using 100 replica variations. In each case, the systematic uncertainty was determined to be negligible.
The systematic uncertainty from the pileup and PDF contributions to the MS DV reconstruction efficiency for signal was evaluated with a procedure similar to that for the trigger, and again the systematic uncertainties were determined to be negligible.
 
The mismodeling of the L1 muon trigger efficiency in MC simulation described in Section~\ref{sec:trigvtxreco} contributes a systematic uncertainty of 20\% (24\%) in the barrel (endcaps). Another source of systematic uncertainty is the uncertainty in the corrected efficiencies that is due to the MC mismodeling in the vertex reconstruction described in Section~\ref{sec:2MSVxEff}. The corrected vertex reconstruction efficiencies were also estimated with a second method, in which events are weighted such that the MC distribution of the number of reconstructed  tracklets matches the data distribution. The efficiency was re-evaluated after reweighting each vertex according to how many tracklets are associated with it. The average efficiency variation is computed and its difference from the variation calculated using the nominal method described in Section~\ref{sec:2MSVxEff} is taken as the systematic uncertainty, which is 11\% in the barrel and 13\% in the endcaps.
 
A systematic uncertainty associated with the efficiency extrapolation method was estimated by comparing the signal efficiency computed using the fully simulated MC samples with the one extrapolated (using toy MC samples) at the same proper lifetime. The difference between the two efficiencies is used as the systematic uncertainty, and it varies from 1.9\% to 30\%, depending on the kinematics of the sample.
For several of the signal samples, two proper lifetime points were fully simulated: one nominal sample and a secondary sample with longer proper lifetime, as described in Section~\ref{sec:datamc}. The secondary sample was used to cross-check the lifetime-extrapolation procedure, and good closure was found.
 
An uncertainty on the NLO-reweighting of the signal samples is obtained by comparing the 125-GeV mediator NLO \MADGRAPH predictions to next-to-next-to-leading-order (NNLO) accuracy in QCD using \POWHEGBOX v2~\cite{Hamilton:2013fea,Hamilton:2015nsa,Alioli:2010xd,Nason:2004rx,Frixione:2007vw}. This results in an additional uncertainty on the signal efficiency ranging from 0.1 to 4\%.
 
The uncertainty in the combined 2015--2018 integrated luminosity is 1.7\% \cite{ATLAS-CONF-2019-021}, obtained using the LUCID-2 detector \cite{LUCID2} for the primary luminosity measurements.
 
The systematic uncertainties described above change the production cross-section limits by about 15\% on average.

\section{Results}
\label{sec:result}
 
In ATLAS Run~2 data, $0.32\pm0.05$ background events are expected for this analysis. Zero events are observed in the signal region.
 
Upper limits on the production cross-section times branching fraction are derived using the CL$_{\textrm{s}}$ prescription \cite{CLs}, implemented with the pyhf~\cite{pyhf, pyhf_joss} package using a profile likelihood function~\cite{Cowan:2010js}. The likelihood includes a Poisson probability term describing the total number of observed events. Background and signal uncertainties are taken into account using Gaussian terms, as commonly done in these cases. Pseudo-experiments which sample the distribution of the profile likelihood ratio are generated to compute the $p$-value and derive the exclusion limits.
 
For scalar boson benchmark samples with $m_{\varPhi}\neq125$~\GeV, upper limits are set on $\sigma \times B$, where $B$ represents the branching fraction for $\varPhi \rightarrow ss$ assuming 100\% branching fraction of $s$ into the heaviest fermion pairs kinematically accessible. As discussed in Section~\ref{sec:models}, the long-lived scalar mainly decays into $b\overline{b}$, except when $m_s>2m_t$ (i.e. for the sample with $m_{\varPhi}=1000$~\GeV\ and $m_s=475$~\GeV) where the dominant decay is into $t\overline{t}$. For scalar boson benchmark samples with $m_{\varPhi}=125$~\GeV, upper limits are set on $\left(\sigma/\sigma_{\textrm{SM}}\right) \times B$, where $\sigma_{\textrm{SM}} = 48.61$~pb~\cite{HiggsXS13TeVHLLHC} is the SM Higgs boson gluon--gluon fusion production cross-section.
 
\begin{figure}[h!]
\centering
\begin{subfigure}[b]{0.48\textwidth}
\includegraphics[width=\textwidth]{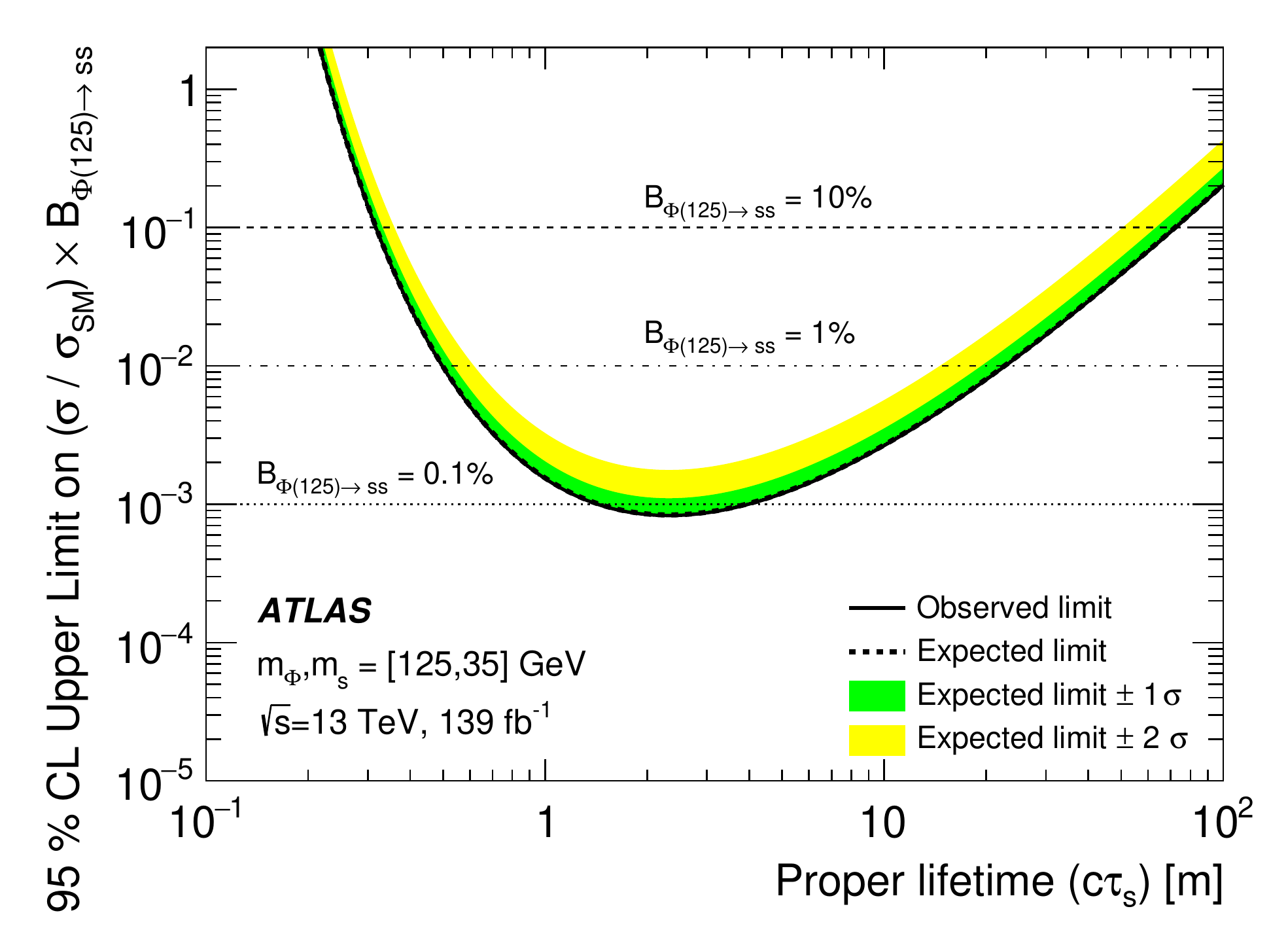}
\caption{}
\end{subfigure}
~
\begin{subfigure}[b]{0.48\textwidth}
\includegraphics[width=\textwidth]{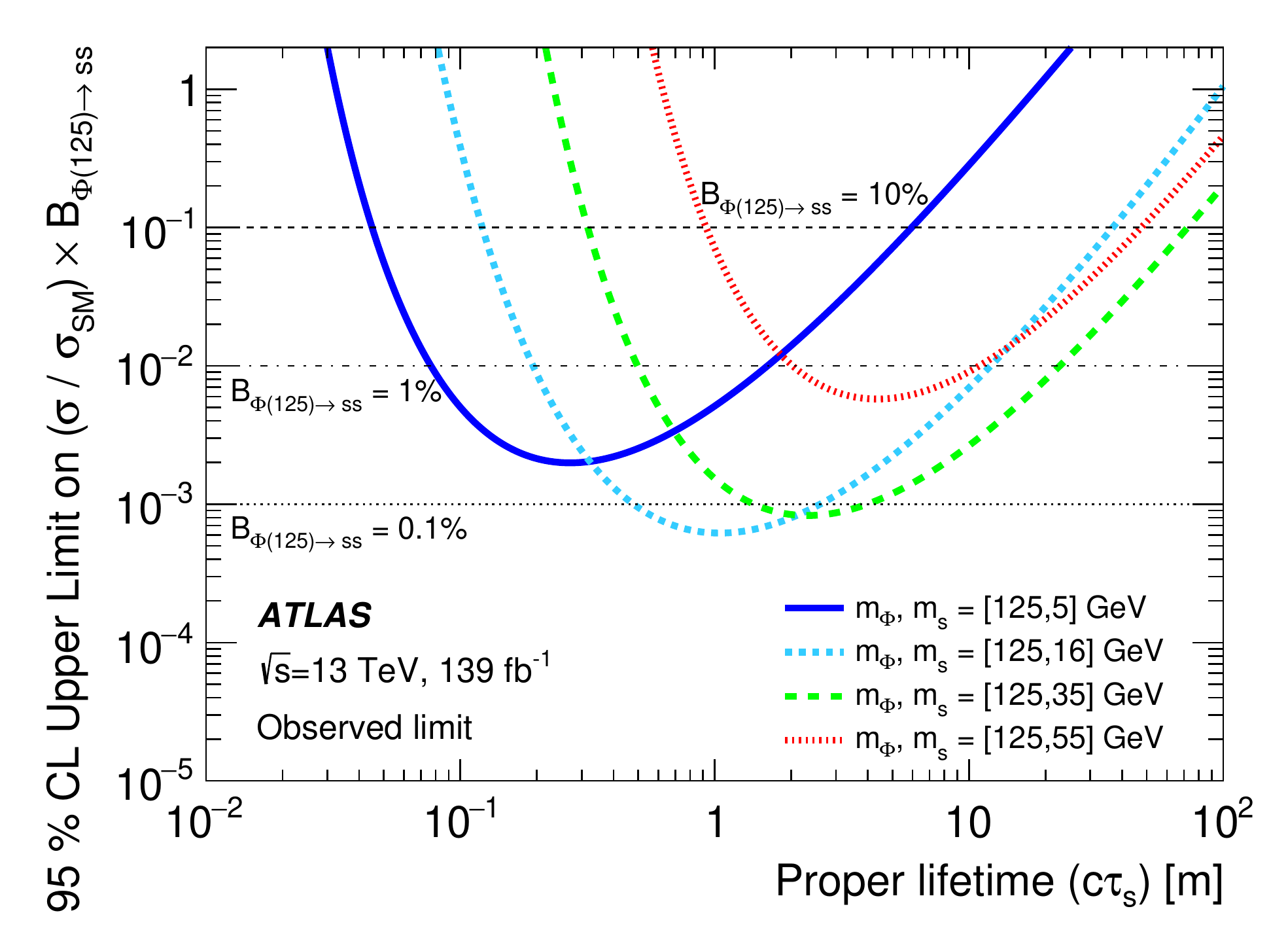}
\caption{}
\end{subfigure} \\
~
\begin{subfigure}[b]{0.48\textwidth}
\includegraphics[width=\textwidth]{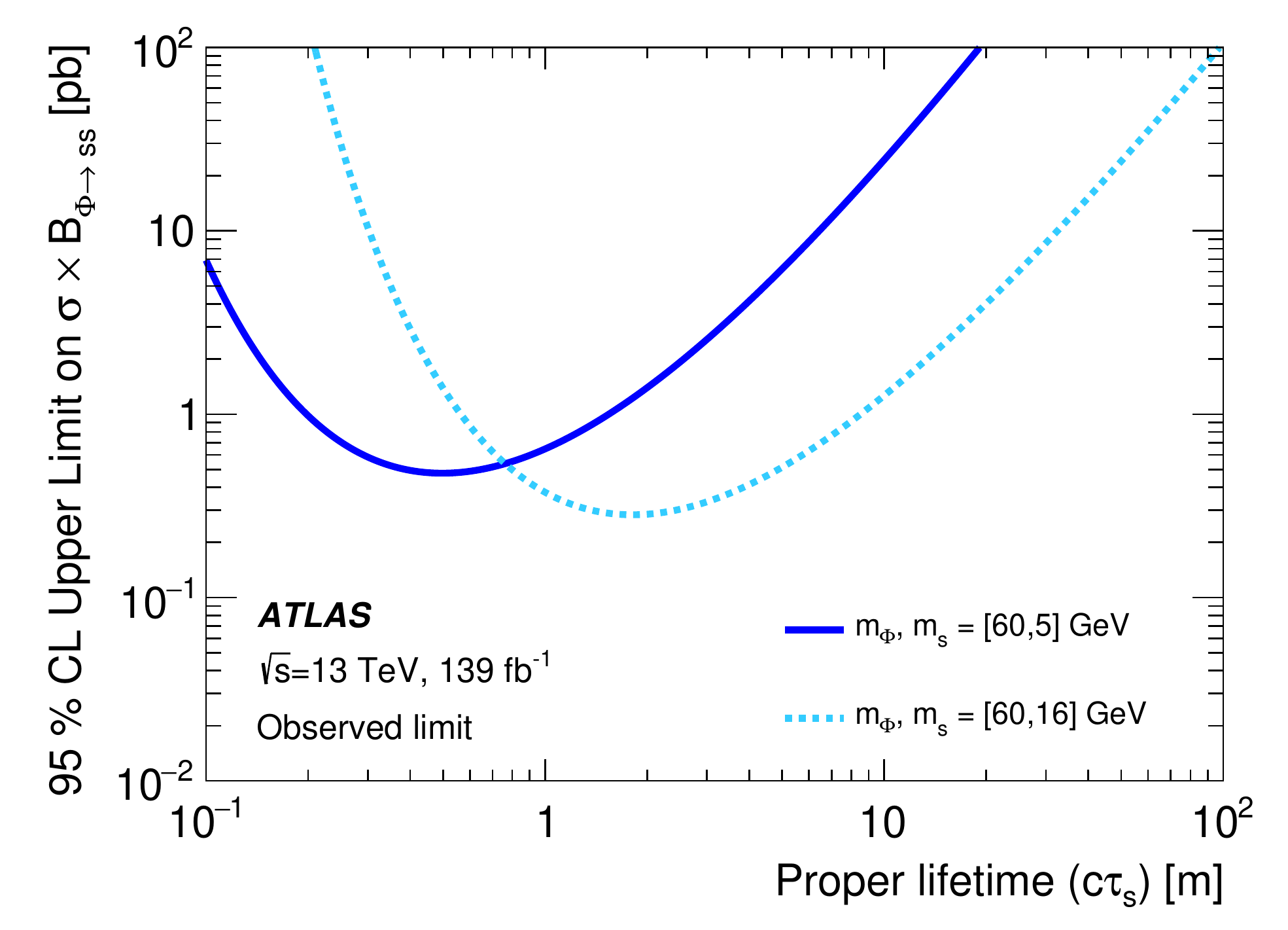}
\caption{}
\end{subfigure}
\begin{subfigure}[b]{0.48\textwidth}
\includegraphics[width=\textwidth]{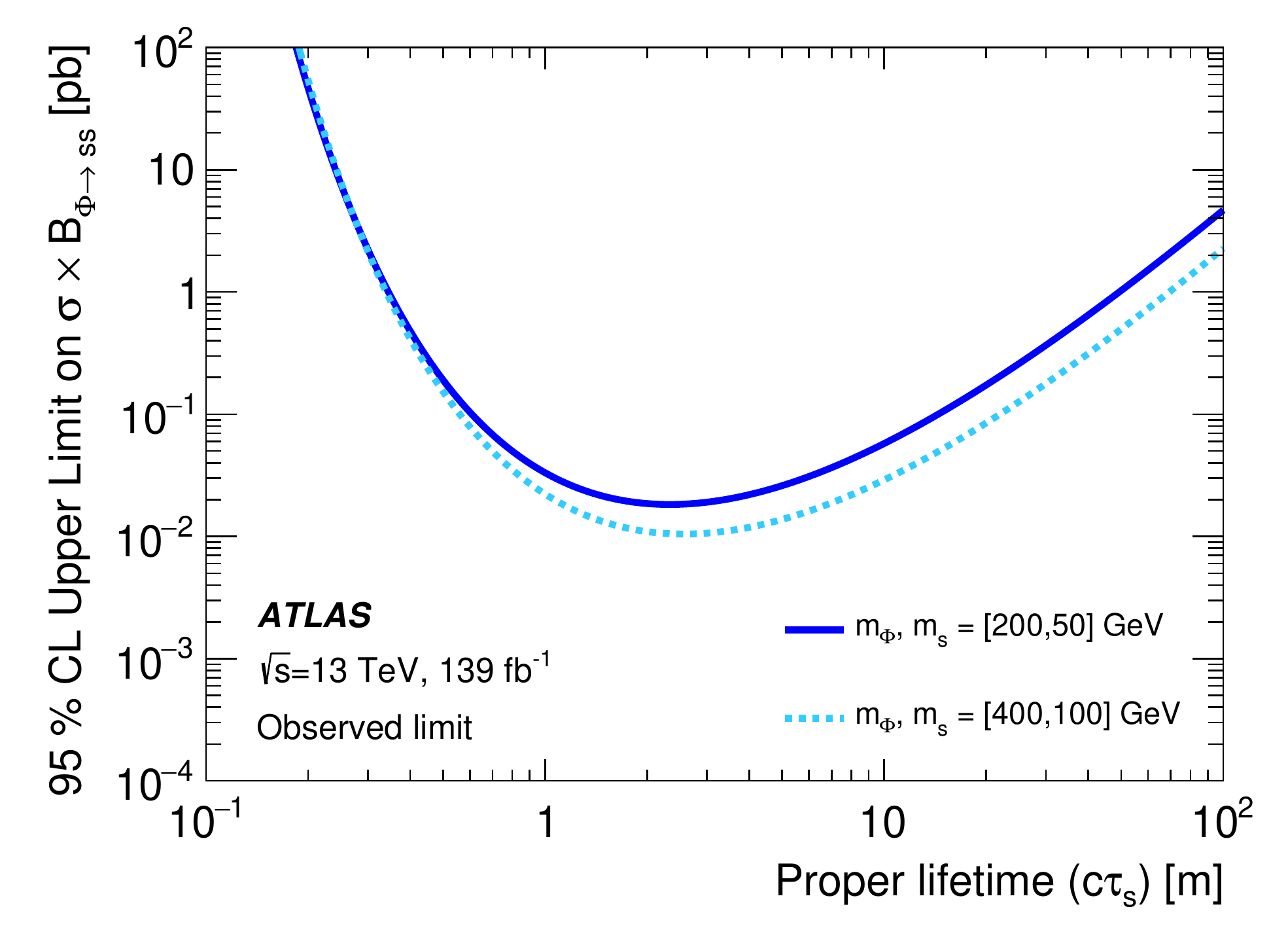}
\caption{}
\end{subfigure}
~
\begin{subfigure}[b]{0.48\textwidth}
\includegraphics[width=\textwidth]{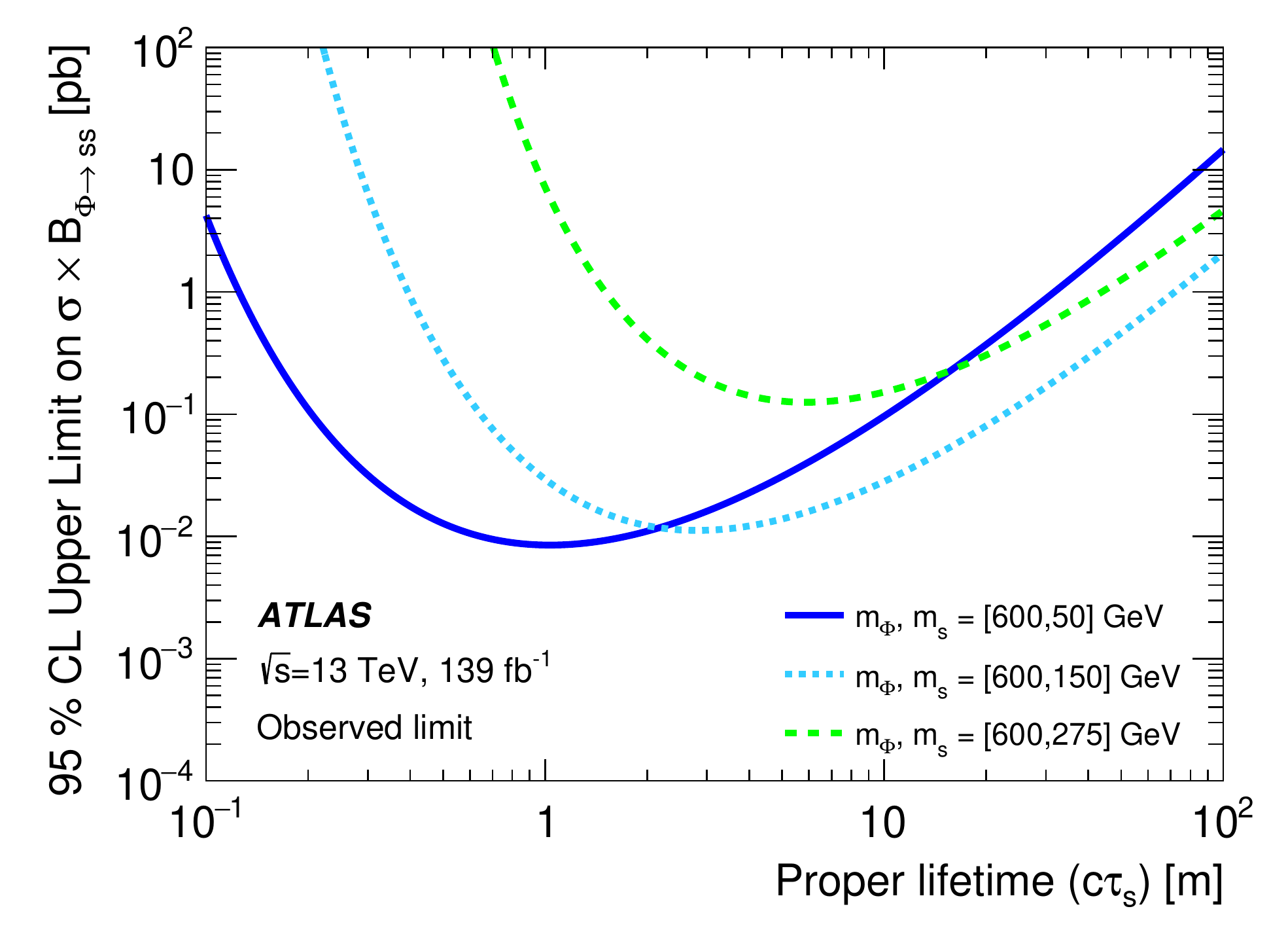}
\caption{}
\end{subfigure}
\begin{subfigure}[b]{0.48\textwidth}
\includegraphics[width=\textwidth]{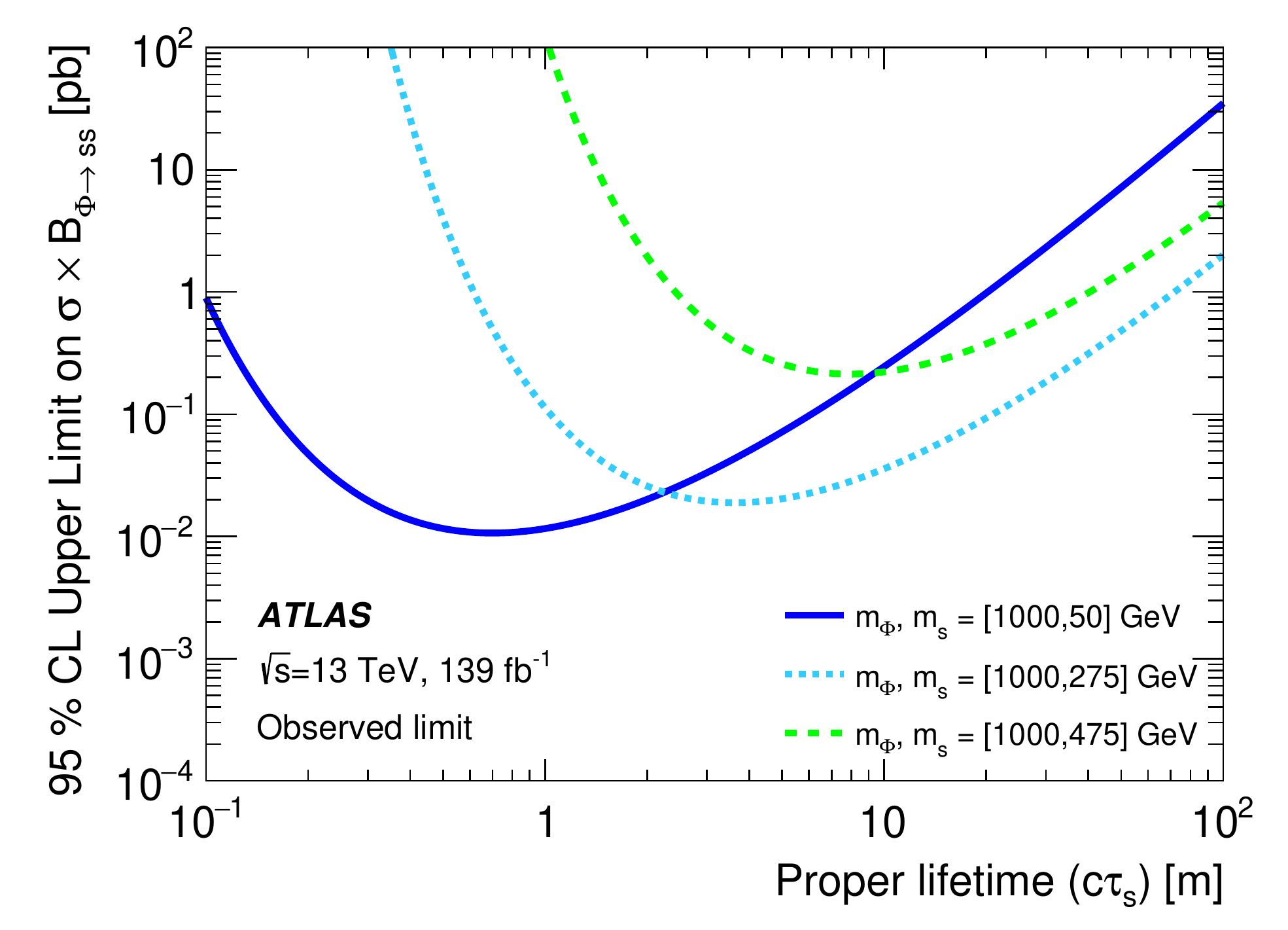}
\caption{}
\end{subfigure}
\caption{Comparison between observed and expected 95\% CL limits (a) on $\left(\sigma/\sigma_{\textrm{SM}}\right) \times B$ for $m_{\varPhi}=125$~\GeV\ and $m_s=35$~\GeV. Observed 95\% CL limits (b) on $\left(\sigma/\sigma_{\textrm{SM}}\right) \times B$ for $m_{\varPhi}=125$~\GeV, and (c)--(f) on $\sigma \times B$ for $m_{\varPhi}\neq125$~\GeV.}
\label{fig:ObsLimits}
\end{figure}

Figure~\ref{fig:ObsLimits} shows a comparison between the expected and observed 95\% CL upper limits for one representative sample as well as the observed 95\% CL limits for all the MC benchmark samples. Observed limits are consistent with the expected ones within the uncertainty. For $m_{\varPhi}=125$~\GeV\ the limits are stronger for intermediate LLP masses, while they become weaker for very low and very high masses. Moreover, the mean proper lifetime $c\tau_s$ at which the upper limit is strongest increases with the long-lived scalar mass. These patterns of behavior are correlated with changes in the trigger and reconstruction efficiencies that depend mainly on the kinematics of the LLP decay.
 
Table~\ref{tab:ExcludedRange_HSS} summarizes the lifetime ranges excluded by the analysis for branching fractions $B(\varPhi(125) \rightarrow ss) = $ 10\%, 1\% and 0.1\% for the scalar boson with $m_\varPhi = 125$~\GeV\ decaying into two long-lived scalars.
 
\begin{table}[htbp!]
\centering
\caption{Ranges of mean proper lifetime excluded at 95\% CL for scalar boson benchmark models with \mbox{$m_\varPhi=125$~\GeV}, assuming a production cross-section times branching fraction equal to 10\%, 1\% and 0.1\% of the SM Higgs boson production cross-section~\cite{HiggsXS13TeVHLLHC}.}\label{tab:ExcludedRange_HSS}
\begin{tabular}{c|cccc}
\toprule
$\varPhi(125) \rightarrow ss   $     &  \multicolumn{3}{c}{Excluded $c\tau_s$ range for $s$ [m]} \\
$m_{s}$~[\GeV]  & $B = 0.1\%$ & $B = 1\%$ & $B = 10\%$  \\
\midrule
~~5  & --         & 0.08--1.6~~  &  0.04--5.9~~ \\
16   & 0.48--2.6  & 0.19--12.2   &  0.12--36.7 \\
35   & ~~1.4--4.0 & 0.49--22.8   &  0.31--72.4 \\
55   & --         & 2.0--11.0    &  0.92--47.6 \\
\bottomrule
\end{tabular}
\end{table}

\clearpage
 
\section{Summary}
\label{sec:summary}
 
This paper presents a search for events with two displaced vertices from pair-produced long-lived particles decaying into hadronic jets using 139~fb$^{-1}$ of \mbox{proton--proton} collisions at $\sqrt{s}=13$~\TeV\ recorded at the LHC by the ATLAS detector during 2015--2018 data-taking period. The search is performed with the same strategy that ATLAS adopted in Run~1 and in a Run~2 search using 2015--2016 data, and benefits from very low background. A data-driven method is used to estimate the expected background in the signal region. No events with two reconstructed displaced vertices in the muon spectrometer are observed in the signal region, and exclusion limits on the LLP production cross-section as a function of its proper lifetime are computed for a scalar portal model where a Higgs boson with a mass of 125~\GeV\ or another short-lived scalar can decay into two long-lived scalars.
For the 125~\GeV\ Higgs boson, the paper reports the first exclusion limits for branching fractions below 0.1\%, while branching fractions above 10\% are excluded at 95\% confidence level for LLP mean proper lifetimes ranging from 4~cm to 72.4~m. In addition, the paper presents the first results for the decay of LLPs into $t\overline{t}$ in the ATLAS muon spectrometer.

\section*{Acknowledgments}

We thank CERN for the very successful operation of the LHC, as well as the
support staff from our institutions without whom ATLAS could not be
operated efficiently.
 
We acknowledge the support of
ANPCyT, Argentina;
YerPhI, Armenia;
ARC, Australia;
BMWFW and FWF, Austria;
ANAS, Azerbaijan;
SSTC, Belarus;
CNPq and FAPESP, Brazil;
NSERC, NRC and CFI, Canada;
CERN;
ANID, Chile;
CAS, MOST and NSFC, China;
Minciencias, Colombia;
MEYS CR, Czech Republic;
DNRF and DNSRC, Denmark;
IN2P3-CNRS and CEA-DRF/IRFU, France;
SRNSFG, Georgia;
BMBF, HGF and MPG, Germany;
GSRI, Greece;
RGC and Hong Kong SAR, China;
ISF and Benoziyo Center, Israel;
INFN, Italy;
MEXT and JSPS, Japan;
CNRST, Morocco;
NWO, Netherlands;
RCN, Norway;
MEiN, Poland;
FCT, Portugal;
MNE/IFA, Romania;
JINR;
MES of Russia and NRC KI, Russian Federation;
MESTD, Serbia;
MSSR, Slovakia;
ARRS and MIZ\v{S}, Slovenia;
DSI/NRF, South Africa;
MICINN, Spain;
SRC and Wallenberg Foundation, Sweden;
SERI, SNSF and Cantons of Bern and Geneva, Switzerland;
MOST, Taiwan;
TENMAK, T\"urkiye;
STFC, United Kingdom;
DOE and NSF, United States of America.
In addition, individual groups and members have received support from
BCKDF, CANARIE, Compute Canada and CRC, Canada;
COST, ERC, ERDF, Horizon 2020 and Marie Sk{\l}odowska-Curie Actions, European Union;
Investissements d'Avenir Labex, Investissements d'Avenir Idex and ANR, France;
DFG and AvH Foundation, Germany;
Herakleitos, Thales and Aristeia programmes co-financed by EU-ESF and the Greek NSRF, Greece;
BSF-NSF and GIF, Israel;
Norwegian Financial Mechanism 2014-2021, Norway;
NCN and NAWA, Poland;
La Caixa Banking Foundation, CERCA Programme Generalitat de Catalunya and PROMETEO and GenT Programmes Generalitat Valenciana, Spain;
G\"{o}ran Gustafssons Stiftelse, Sweden;
The Royal Society and Leverhulme Trust, United Kingdom.
 
The crucial computing support from all WLCG partners is acknowledged gratefully, in particular from CERN, the ATLAS Tier-1 facilities at TRIUMF (Canada), NDGF (Denmark, Norway, Sweden), CC-IN2P3 (France), KIT/GridKA (Germany), INFN-CNAF (Italy), NL-T1 (Netherlands), PIC (Spain), ASGC (Taiwan), RAL (UK) and BNL (USA), the Tier-2 facilities worldwide and large non-WLCG resource providers. Major contributors of computing resources are listed in Ref.~\cite{ATL-SOFT-PUB-2021-003}.
 

\printbibliography
 
 
\begin{flushleft}
\hypersetup{urlcolor=black}
{\Large The ATLAS Collaboration}

\bigskip

\AtlasOrcid[0000-0002-6665-4934]{G.~Aad}$^\textrm{\scriptsize 98}$,    
\AtlasOrcid[0000-0002-5888-2734]{B.~Abbott}$^\textrm{\scriptsize 124}$,    
\AtlasOrcid[0000-0002-7248-3203]{D.C.~Abbott}$^\textrm{\scriptsize 99}$,    
\AtlasOrcid[0000-0002-2788-3822]{A.~Abed~Abud}$^\textrm{\scriptsize 34}$,    
\AtlasOrcid[0000-0002-1002-1652]{K.~Abeling}$^\textrm{\scriptsize 51}$,    
\AtlasOrcid[0000-0002-2987-4006]{D.K.~Abhayasinghe}$^\textrm{\scriptsize 91}$,    
\AtlasOrcid[0000-0002-8496-9294]{S.H.~Abidi}$^\textrm{\scriptsize 27}$,    
\AtlasOrcid[0000-0002-9987-2292]{A.~Aboulhorma}$^\textrm{\scriptsize 33e}$,    
\AtlasOrcid[0000-0001-5329-6640]{H.~Abramowicz}$^\textrm{\scriptsize 157}$,    
\AtlasOrcid[0000-0002-1599-2896]{H.~Abreu}$^\textrm{\scriptsize 156}$
% ...
\bigskip
\\

% ...
$^{27}$Physics Department, Brookhaven National Laboratory, Upton NY; United States of America.\\
% ...
$^{33}$$^{(a)}$Facult\'e des Sciences Ain Chock, R\'eseau Universitaire de Physique des Hautes Energies - Universit\'e Hassan II, Casablanca;$^{(b)}$Facult\'{e} des Sciences, Universit\'{e} Ibn-Tofail, K\'{e}nitra;$^{(c)}$Facult\'e des Sciences Semlalia, Universit\'e Cadi Ayyad, LPHEA-Marrakech;$^{(d)}$LPMR, Facult\'e des Sciences, Universit\'e Mohamed Premier, Oujda;$^{(e)}$Facult\'e des sciences, Universit\'e Mohammed V, Rabat;$^{(f)}$Mohammed VI Polytechnic University, Ben Guerir; Morocco.\\
$^{34}$CERN, Geneva; Switzerland.\\
% ...
$^{51}$II. Physikalisches Institut, Georg-August-Universit\"{a}t G\"ottingen, G\"ottingen; Germany.\\
% ...
$^{91}$Department of Physics, Royal Holloway University of London, Egham; United Kingdom.\\
% ...
$^{98}$CPPM, Aix-Marseille Universit\'e, CNRS/IN2P3, Marseille; France.\\
$^{99}$Department of Physics, University of Massachusetts, Amherst MA; United States of America.\\
% ...
$^{124}$Homer L. Dodge Department of Physics and Astronomy, University of Oklahoma, Norman OK; United States of America.\\
% ...
$^{156}$Department of Physics, Technion, Israel Institute of Technology, Haifa; Israel.\\
$^{157}$Raymond and Beverly Sackler School of Physics and Astronomy, Tel Aviv University, Tel Aviv; Israel.\\
% ...
$^{a}$ Also at Borough of Manhattan Community College, City University of New York, New York NY; United States of America.\\
$^{b}$ Also at Bruno Kessler Foundation, Trento; Italy.\\
$^{c}$ Also at Center for High Energy Physics, Peking University; China.\\
$^{d}$ Also at Centro Studi e Ricerche Enrico Fermi; Italy.\\
$^{e}$ Also at CERN, Geneva; Switzerland.\\
$^{f}$ Also at D\'epartement de Physique Nucl\'eaire et Corpusculaire, Universit\'e de Gen\`eve, Gen\`eve; Switzerland.\\
% ...

\end{flushleft}


\clearpage

\end{document}